\documentclass[fleqn,usenatbib]{mnras}
\usepackage{pdflscape} 
\usepackage{longtable}

\usepackage{verbatim}
\usepackage{hyperref}
\hypersetup{
    colorlinks=true,
    linkcolor=blue,
    filecolor=magenta,      
    urlcolor=cyan,
    citecolor=blue,
}

\usepackage{etoolbox,xparse}
\usepackage[T2A,T1]{fontenc}
\usepackage[utf8]{inputenc}
\usepackage[bulgarian,british]{babel}
\usepackage{datetime2}
\usepackage[usenames,dvipsnames]{xcolor}
\usepackage{rotating}

\usepackage{multirow}
\usepackage{booktabs}  
\usepackage{graphicx}
\usepackage{amsmath}
\usepackage{newtxtext}
\usepackage{newtxmath}
\usepackage[style=english,english=british]{csquotes}  
\usepackage{siunitx}  
\DeclareSIPostPower{\nominal}{N}
\DeclareSIQualifier{\earth}{\ensuremath{\oplus}}
\DeclareSIQualifier{\jupiter}{J}
\DeclareSIQualifier{\planet}{p}
\DeclareSIQualifier{\etoile}{\ensuremath{\star}}
\DeclareSIQualifier{\sun}{\ensuremath{\odot}}
\DeclareSIUnit\angstrom{\AA}
\DeclareSIUnit{\au}{au}
\DeclareSIUnit{\density}{\ensuremath{\mathnormal{\rho}}}
\DeclareSIUnit{\erg}{erg}
\DeclareSIUnit{\luminosity}{\ensuremath{\mathrm{L}}}
\DeclareSIUnit{\magnitude}{mag}
\DeclareSIUnit{\mass}{\ensuremath{\mathrm{M}}}
\DeclareSIUnit{\mas}{\milliarcsecond}
\DeclareSIUnit{\milliarcsecond}{mas}
\DeclareSIUnit{\parsec}{pc}
\DeclareSIUnit{\radius}{\ensuremath{\mathrm{R}}}
\DeclareSIUnit{\year}{yr}
\usepackage{xcolor}

\newcommand{\teff}{\ensuremath{T_{\rm eff}}}

\newcommand{\logg}{\ensuremath{\log g}}
\newcommand{\feh}{[Fe/H]}

\newcommand{\kepler}{{\it Kepler}}

\newcommand{\tess}{{\it TESS}}

\newcommand{\angstrom}{\textup{\AA}}

\title[Warm Sub-Saturn Eccentricities in \tess]{The Eccentricity Distribution of Warm Sub-Saturns in \tess}

\author[T.R. Fairnington et al.]{%
Tyler~R.~Fairnington
\textsuperscript{\href{https://orcid.org/0000-0002-0692-7822}{\includegraphics[width=2.5mm]{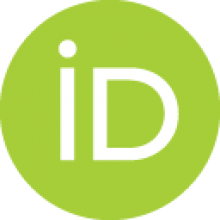}}}%
,\textsuperscript{1}%
\thanks{Email: tyler.fairnington@usq.edu.au}
Jiayin~Dong%
\textsuperscript{\href{https://orcid.org/}{\includegraphics[width=2.5mm]{orcid-ID.png}}}%
,\textsuperscript{2, 3}
\thanks{Flatiron Research Fellow}
Chelsea~X.~Huang%
\textsuperscript{\href{https://orcid.org/0000-0003-0918-7484}{\includegraphics[width=2.5mm]{orcid-ID.png}}}%
,\textsuperscript{1}
\thanks{ARC Future Fellow}
Emma Nabbie%
\textsuperscript{\href{https://orcid.org/0000-0003-0571-2245}{\includegraphics[width=2.5mm]{orcid-ID.png}}}%
,\textsuperscript{1}
George Zhou%
\textsuperscript{\href{https://orcid.org/}{\includegraphics[width=2.5mm]{orcid-ID.png}}}%
,\textsuperscript{1}
\thanks{ARC Future Fellow}
\newauthor
Duncan Wright,\textsuperscript{1}%
Karen A.\ Collins%
\textsuperscript{\href{https://orcid.org/0000-0001-6588-9574}{\includegraphics[width=2.5mm]{orcid-ID.png}}}%
,\textsuperscript{4}
Jon M. Jenkins
\textsuperscript{\href{https://orcid.org/0000-0002-4715-9460}{\includegraphics[width=2.5mm]{orcid-ID.png}}}%
,\textsuperscript{5}
David W. Latham
\textsuperscript{\href{https://orcid.org/0000-0001-9911-7388}{\includegraphics[width=2.5mm]{orcid-ID.png}}}%
,\textsuperscript{4}
George Ricker
\textsuperscript{\href{https://orcid.org/0000-0003-2058-6662}{\includegraphics[width=2.5mm]{orcid-ID.png}}}%
,\textsuperscript{6}
\newauthor
Samuel N. Quinn%
\textsuperscript{\href{https://orcid.org/0000-0002-8964-8377}{\includegraphics[width=2.5mm]{orcid-ID.png}}}%
,\textsuperscript{4}
Sara Seager
\textsuperscript{\href{https://orcid.org/0000-0002-6892-6948}{\includegraphics[width=2.5mm]{orcid-ID.png}}}%
,\textsuperscript{7, 6, 8}
Avi Shporer
,\textsuperscript{6}
Roland Vanderspek
\textsuperscript{\href{https://orcid.org/0000-0001-6763-6562}{\includegraphics[width=2.5mm]{orcid-ID.png}}}%
,\textsuperscript{6}
Joshua N. Winn
\textsuperscript{\href{https://orcid.org/0000-0002-4265-047X}{\includegraphics[width=2.5mm]{orcid-ID.png}}}%
,\textsuperscript{9}
\newauthor
Calvin Ajizian
\textsuperscript{\href{https://orcid.org/0009-0008-8562-108X}{\includegraphics[width=2.5mm]{orcid-ID.png}}}%
,\textsuperscript{10}
Akihiko Fukui
\textsuperscript{\href{https://orcid.org/0000-0002-4909-5763}{\includegraphics[width=2.5mm]{orcid-ID.png}}}%
,\textsuperscript{11, 12}
David Baker
\textsuperscript{\href{https://orcid.org/0000-0002-2970-0532}{\includegraphics[width=2.5mm]{orcid-ID.png}}}%
,\textsuperscript{13}
Giuseppe Conzo %
\textsuperscript{\href{https://orcid.org/0000-0002-2412-1558}{\includegraphics[width=2.5mm]{orcid-ID.png}}}%
,\textsuperscript{14}
Robert Scott Fisher %
\textsuperscript{\href{https://orcid.org/0000-0002-9643-7543}{\includegraphics[width=2.5mm]{orcid-ID.png}}}%
,\textsuperscript{10}
\newauthor
Raquel Forés-Toribio %
\textsuperscript{\href{https://orcid.org/0000-0002-6482-2180}{\includegraphics[width=2.5mm]{orcid-ID.png}}}%
,\textsuperscript{15, 16}
Tianjun Gan %
\textsuperscript{\href{https://orcid.org/0000-0002-4503-9705}{\includegraphics[width=2.5mm]{orcid-ID.png}}}%
,\textsuperscript{17}
Alexey Garmash %
\textsuperscript{\href{https://orcid.org/0000-0003-2599-1405}{\includegraphics[width=2.5mm]{orcid-ID.png}}}%
,\textsuperscript{18}
Kai Ikuta
\textsuperscript{\href{https://orcid.org/0000-0002-5978-057X}{\includegraphics[width=2.5mm]{orcid-ID.png}}}%
,\textsuperscript{19}
Adam Lark %
,\textsuperscript{20}
\newauthor
Jerome P. de Leon
\textsuperscript{\href{https://orcid.org/0000-0002-6424-3410}{\includegraphics[width=2.5mm]{orcid-ID.png}}}%
,\textsuperscript{19}
Katherine Linnenkohl %
\textsuperscript{\href{https://orcid.org/0009-0009-7566-8420}{\includegraphics[width=2.5mm]{orcid-ID.png}}}%
,\textsuperscript{10}
Christopher R. Mann %
\textsuperscript{\href{https://orcid.org/0000-0002-9312-0073}{\includegraphics[width=2.5mm]{orcid-ID.png}}}%
,\textsuperscript{21}
Owen Mitchem %
\textsuperscript{\href{https://orcid.org/0009-0005-5648-7107}{\includegraphics[width=2.5mm]{orcid-ID.png}}}%
,\textsuperscript{10}
\newauthor
Mayuko Mori
\textsuperscript{\href{https://orcid.org/0000-0003-1368-6593}{\includegraphics[width=2.5mm]{orcid-ID.png}}}%
,\textsuperscript{22, 23}
Jose A. Muñoz %
\textsuperscript{\href{https://orcid.org/0000-0001-9833-2959}{\includegraphics[width=2.5mm]{orcid-ID.png}}}%
,\textsuperscript{15, 16}
Norio Narita
\textsuperscript{\href{https://orcid.org/0000-0001-8511-2981}{\includegraphics[width=2.5mm]{orcid-ID.png}}}%
,\textsuperscript{11, 22, 12}
Adam Popowicz %
\textsuperscript{\href{https://orcid.org/0000-0003-3184-5228}{\includegraphics[width=2.5mm]{orcid-ID.png}}}%
,\textsuperscript{24}
\newauthor
Don Radford %
\textsuperscript{\href{https://orcid.org/0000-0002-3940-2360}{\includegraphics[width=2.5mm]{orcid-ID.png}}}%
,\textsuperscript{25}
Justus Randolph
,\textsuperscript{26}
Fabian Rodriguez Frustaglia %
,\textsuperscript{26}
Richard P. Schwarz 
\textsuperscript{\href{https://orcid.org/0000-0001-8227-1020}{\includegraphics[width=2.5mm]{orcid-ID.png}}}%
,\textsuperscript{4}
\newauthor
Chris Stockdale %
\textsuperscript{\href{https://orcid.org/0000-0003-2163-1437}{\includegraphics[width=2.5mm]{orcid-ID.png}}}%
,\textsuperscript{27}
Jiaqi Wang
\textsuperscript{\href{https://orcid.org/0000-0001-8665-4598}{\includegraphics[width=2.5mm]{orcid-ID.png}}}%
,\textsuperscript{28}
Noriharu Watanabe
\textsuperscript{\href{https://orcid.org/0000-0002-7522-8195}{\includegraphics[width=2.5mm]{orcid-ID.png}}}%
,\textsuperscript{19}
Francis P. Wilkin
\textsuperscript{\href{https://orcid.org/0000-0003-2127-8952}{\includegraphics[width=2.5mm]{orcid-ID.png}}}%
,\textsuperscript{29}
\newauthor
Krzysztof Sz. Zieliński
\textsuperscript{\href{https://orcid.org/0009-0001-0389-8907}{\includegraphics[width=2.5mm]{orcid-ID.png}}}%
,\textsuperscript{30, 31}
Emma Esparza-Borges
\textsuperscript{\href{https://orcid.org/0000-0002-2341-3233}{\includegraphics[width=2.5mm]{orcid-ID.png}}}%
,\textsuperscript{12, 32}
Felipe Murgas
\textsuperscript{\href{https://orcid.org/0000-0001-9087-1245}{\includegraphics[width=2.5mm]{orcid-ID.png}}}%
,\textsuperscript{12, 32}
Enric Pall\'e
\textsuperscript{\href{https://orcid.org/0000-0003-0987-1593}{\includegraphics[width=2.5mm]{orcid-ID.png}}}%
,\textsuperscript{12, 32}
\newauthor
Hannu Parviainen \textsuperscript{\href{https://orcid.org/0000-0001-5519-1391}{\includegraphics[width=2.5mm]{orcid-ID.png}}}%
,\textsuperscript{12, 32}
Sel\c{c}uk Yal\c{c}{\i}nkaya \textsuperscript{\href{https://orcid.org/0000-0002-5224-247X}{\includegraphics[width=2.5mm]{orcid-ID.png}}}%
,\textsuperscript{33, 34}
Ozg\"ur Ba\c{s}t\"urk \textsuperscript{\href{https://orcid.org/0000-0002-4746-0181}{\includegraphics[width=2.5mm]{orcid-ID.png}}}%
,\textsuperscript{33, 34}
\\
Affiliations are listed at the end of the paper
}

\begin{document}
\label{firstpage}
\pagerange{\pageref{firstpage}--\pageref{lastpage}}
\maketitle

\begin{abstract}
We present the eccentricity distribution of warm sub-Saturns (4--8 $R_{\oplus}$, 8--200 day periods) as derived from an analysis of transit light curves from NASA's \textit{Transiting Exoplanet Survey Satellite} (\tess) mission.
We use the ``photoeccentric'' effect to constrain the eccentricities of {76} planets, {comprising 60 and 16 from single- and multi-transiting systems, respectively.}
We employ Hierarchical Bayesian Modelling to infer the eccentricity distribution of the population, testing both a Beta and Mixture Beta distribution. 
We identify a few highly eccentric (${e\sim0.7-0.8}$) warm sub-Saturns with eccentricities that appear too high to be explained by disk migration or planet-planet scattering alone, suggesting high-eccentricity migration may play a role in their formation. {The majority of the population have a mean eccentricity of {$\bar{e} = 0.103^{+0.047}_{-0.045}$}, consistent with both planet-disk and planet-planet interactions.}
Notably, we find that the highly eccentric sub-Saturns occur in single-transiting systems. 
This study presents the first evidence {at the population level} that the eccentricities of sub-Saturns may be sculpted by dynamical processes. 

\end{abstract}


\section{Introduction}
\label{sec:introduction}

Sub-Saturns (4--8 $R_{\oplus}$) are a puzzling planet class due to their unique observed properties. Unlike the general planet population, they do not exhibit strong correlations between mass and radius \citep{Petigura:2017}. They are also intrinsically rare, occurring around only $\sim$3.1\% of FGK stars within 100 days \citep{Kunimoto:2020}. Multiple formation and evolution mechanisms are suggested to contribute to the observed properties of sub-Saturns, including ``failed gas giants" from runaway accretion \citep{Lee:2016}, ``boil-off gas giants" due to photoevaporation \citep{Owen:2016, Hallatt:2022}, tidally inflated sub-Neptunes \citep{Millholland:2020}, and planet-planet merger products \citep{Ginzburg:2020}.

Orbital eccentricities may offer insights into whether single or multiple formation and evolution channels contribute to the sub-Saturn population. The eccentricity of a planet serves as a key indicator of its formation and evolution, which has been used to show that planet-planet interactions sculpt planet populations \citep{Ford:2008, Juric:2008, Xie:2016, VanEylen:2019, Dong:2021}. A study by \citet{Petigura:2017} examined 23 sub-Saturns and showed tentative evidence that some single sub-Saturns possess modest eccentricities, suggesting a role for planet-planet mergers in their formation. 
\citet{Nowak:2020} later reported similar findings, including the discovery of a moderately eccentric sub-Saturn ($e\sim0.35$). However, these studies primarily focused on shorter orbital periods where planet-star tidal interactions may have dampened their eccentricities \citep{Goldreich:1966}.

The discovery of Kepler-1656\,b \citep{Brady:2018, angelo:2022} has highlighted the importance of studying warm sub-Saturns (8-200 day periods). As the only confirmed highly eccentric warm sub-Saturn ($e\approx0.84$), Kepler-1656\,b may be the outcome of both a merger \citep{Brady:2018, Millholland:2020} and subsequent high eccentricity migration (HEM), as merger scenarios alone are unable to excite such high eccentricities. The detection of an exterior non-transiting Jovian companion further supports this interpretation \citep{angelo:2022}. However, whether Kepler-1656\,b represents a unique case or indicates a broader population of significantly eccentric sub-Saturns remains an open question. 

The \textit{Transiting Exoplanet Survey Satellite} (\tess, \citealt{Ricker:2015}) provides an opportunity to address this question. As an all-sky survey, \tess{} is capable of finding sub-Saturns around some of the brightest stars in the sky \citep{Kipping:2019, Jenkins:2020, Fairnington:2024, Barber:2024} and has already discovered various warm sub-Saturn systems \citep{Newton:2019, Dransfield:2022, Kunimoto:2023, Dai:2023, Polanski:2024}. This expanded sample enables the first systematic study of warm sub-Saturn eccentricities and their potential formation pathways.

In this paper, we examine the eccentricity distribution of warm sub-Saturns observed by \tess{} and explore the implications of the observed distribution. In Section \S\ref{sec:data}, we describe our sample selection and light curve processing. Section \S\ref{sec:analysis} presents the stellar and light curve modelling of each candidate system. In Section \S\ref{sec:distribution}, we infer the eccentricity distribution for the warm sub-Saturn population. Finally, in Section \S\ref{sec:discussion} we present our findings in the context of the planet population, discussing potential formation pathways for sub-Saturns.

\section{Data}
\label{sec:data}
\subsection{Warm Sub-Saturn Candidates in \tess{}}

We compile an initial list of \tess{} warm sub-Saturn  candidates with orbital periods between 8 and 200 days. The minimum period threshold corresponds to the approximate period that tidal circularization is weak enough to result in primordially eccentric planets retaining some eccentricity, and matches that of a previous study of warm Jupiters in \tess{}  \citep{Dong:2021}. The candidates are selected from the \tess{} Object of Interest (TOI) catalog\footnote{The TOI catalog was downloaded from NASA Exoplanet Archive: https://exoplanetarchive.ipac.caltech.edu/cgi-bin/TblView/nph-tblView?app=ExoTbls\&config=TOI \citep{exoplanetarchive}} \citep{TOI} as of January 29, 2024, where we only include candidates identified within the first five years of \tess{} (Sectors 1-69). We also applied a magnitude threshold to only include planets around stars brighter than $T_{mag}$ of 12 so that the transits of the sub-Saturns will be detected with a relatively high signal-to-noise ratio. We exclude candidates labelled as False Positive (FP), False Alarm (FA), or Ambiguous Planet Candidate (APC) by both the \tess{} team and the Follow-up Observing Program (TFOP, \citealt{TFOP}). These criteria yield an initial catalog of {132} warm sub-Saturn candidates.  
We further separated the candidates into two categories: those with only one transiting planetary candidate identified in the system (singles) and those in systems hosting multiple transiting candidates (multis). This provides an initial pool of {100} single transiting warm sub-Saturn candidates \footnote{{Kepler-396 (TOI-4579) and Kepler-450 (TOI-5993)} only have one transiting planet identified by \tess{} data, however they have additional small planets confirmed by \kepler{}, and as such they are included in the multi-transiting sample in our analysis.} and 32 sub-Saturns in multi-transiting systems. 

\begin{figure}
\includegraphics[width=\linewidth]{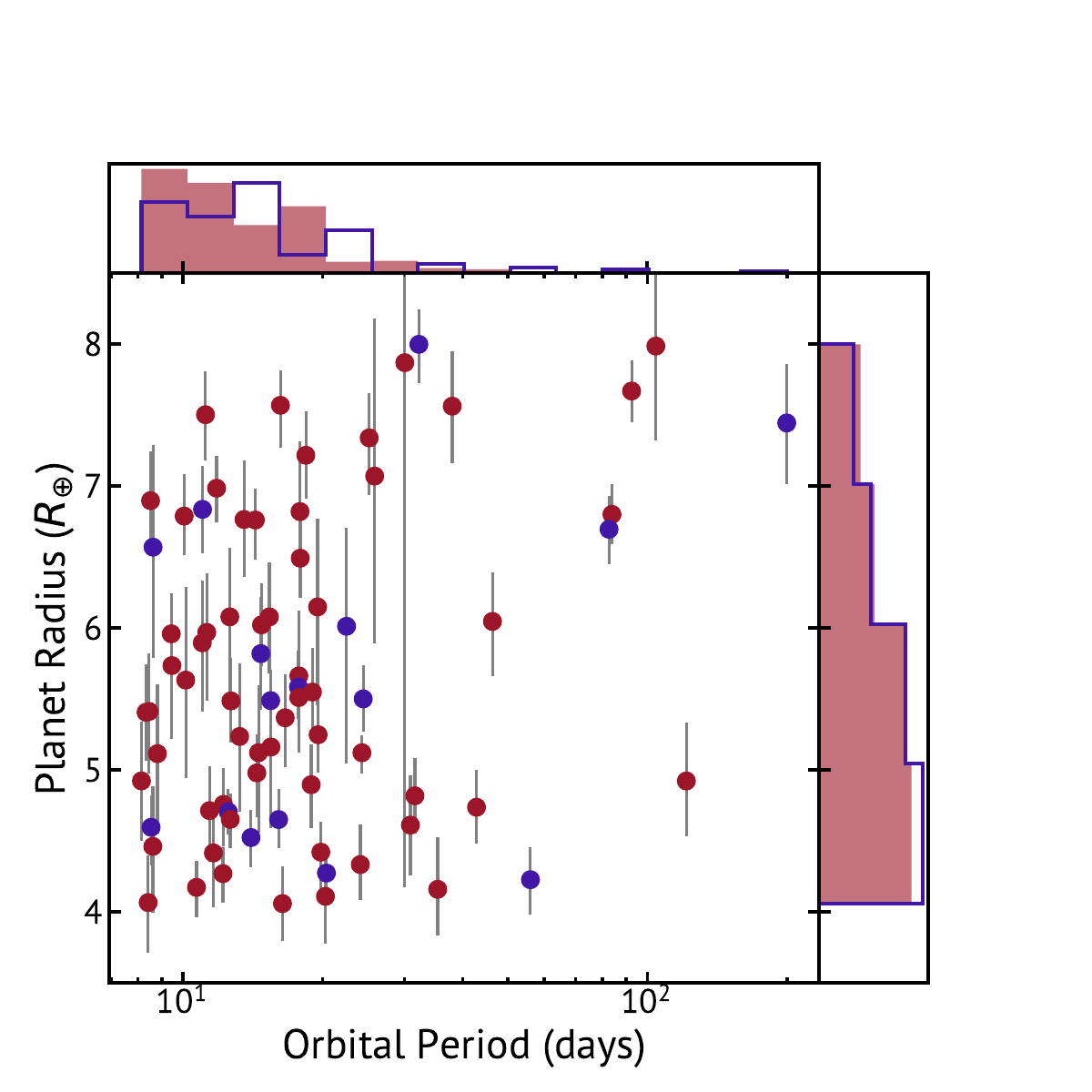}
\caption{The period-radius of the distribution of sub-Saturns in our sample used for the fiducial HBM model. Red points are sub-Saturns in single-transiting systems, dark purple points are sub-Saturns in multi-transiting systems. The histograms represent the period-radius data as probability densities. \label{fig:period-radius}}
\end{figure}

We initially ruled out 4 candidates with no \textit{Gaia} DR2  \citep{gaiadr2} stellar density constraints, as well as the circumbinary planet TOI-1338AB b \citep{toi1338_1}, as both compromise assumptions made about the stellar density required for the ``photoeccentric'' effect (see Section \S\ref{sec:analysis}). We also make a series of cuts to the initial sample following light curve modelling as described in Section \S\ref{sec:analysis}. The period-radius distribution of the final sample can be seen in Figure \ref{fig:period-radius}. 
{
\begin{table}[ht]
\centering
\caption{The sample selection process for our catalog involving the selection step and remaining sample.}
\label{tab:sample_selection}
\begin{tabular}{@{} l S[table-format=2.0] S[table-format=3.0] @{}}
\toprule
{Selection Step} & {Sample Remaining} \\ 
\midrule
Initial Catalog           & 132 \\ 
Missing Gaia DR2 Stellar Density  &  127\\ 
Unresolved Period Alias   &  116 \\ 
Impact Parameter $\ge$ 0.9  & 101 \\ 
Planet Radius outside 4--8 $R_{\oplus}$  & 76 \\ 
\bottomrule
\end{tabular}
\end{table}
}

\subsection{Processing Light Curves}
\label{ssec:modelling}

 We retrieve the \tess{} light curves from the publicly available Mikulski Archive for Space Telescopes (MAST) using the \textsc{lightkurve} Python package \citep{lightkurve}. We download all available sectors up to sector 69, prioritizing the shortest-cadence data when multiple-cadence observations are available for a sector. We use the Simple Aperture Photometry (SAP) flux \citep{SPOC, twicken:PA2010SPIE, morris:PA2020KDPH, QLP} for our analysis. We take into account contamination from neighbouring stars in the aperture when using SPOC SAP data \footnote{QLP SAP FLUX is already decontaminated.}. For all sectors of \tess{} data, quality flags are applied to exclude cadences affected by scattered light or other systematics. In some QLP sectors, the flux error is unavailable, we estimate the flux error for each sector by approximating it as 1.48$\times$ the median absolute deviation (MAD) of the raw flux. We detrend each systems' light curves simultaneously to our global modelling, as described by Section \S\ref{sec:analysis}.

\section{Analysis}
\label{sec:analysis}
\subsection{Stellar Parameters}
\label{ssec:stellarmodelling}

We use \textsc{astroARIADNE} \citep{ariadne} to perform an independent stellar spectral energy distribution (SED) modelling, so that we obtain uniformly derived priors for the true stellar density to constrain the planets' eccentricities. 

\textsc{astroARIADNE} employs a Bayesian Model Averaging (BMA) approach, which fits multiple models to account for model-specific systematic biases. We use the following models in our analysis: Phoenixv2 \citep{Husser:2013},  BT-Settl \citep{Allard:2012}, BT-NextGen \citep{Allard:2012}, BT-Cond \citep{Hauschildt:1999,Allard:2012}, and Kurucz93 \citep{Kurucz:1993}. Each model is individually weighted, resulting in a combined posterior probability, weighted by the likelihood of each model. 
The result is then used to interpret the \textsc{MIST} isochrone \citep{MIST} to obtain the final stellar parameters.   
For the SED fitting, we used Tycho-2 \citep{Tycho2} B and V, Gaia DR3 \citep{fabricius2021} G, Bp and Rp, Two-Micron All-Sky Survey (2MASS) \citep{2MASS} J, H, K, and Wide-field Infrared Survey Explorer (WISE) \citep{WISE} $W_1$ and $W_2$ bands, with an uncertainty floor applied to all bands based on \citet{Eastman:2019}. In instances where no Tycho-2 band observations have been conducted for a target, we use the Johnson B and V magnitudes, but cut these magnitudes if they differ from the Gaia G band by more than 3 magnitudes, as indicative of an incorrect measurement. We use Gaia DR3 parallax measurements \citep{fabricius2021} as a prior on the distance, and a prior on extinction to constrain an upper limit using the galactic dust maps from \citet{Avdustmap}. {We utilize the default priors on stellar effective temperature (\teff), metallicity (\feh) and log surface gravity (\logg), which places empirical priors drawn from the RAVE survey's \teff, \feh, and \logg{} distributions \citep{RAVE}}. 

\begin{figure}
\includegraphics[width=\linewidth]{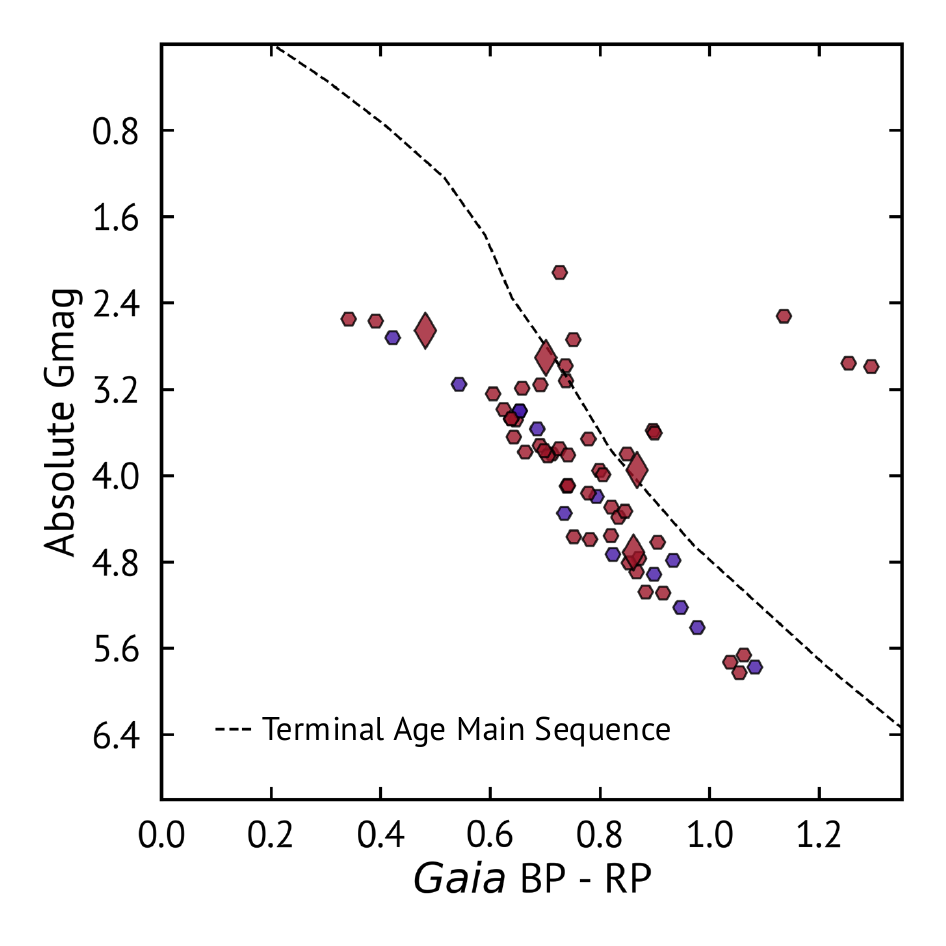}
\caption{Hertzsprung–Russell diagram of the host stars of our final sample. Red points indicate single candidate hosts, while dark purple represents multi candidate hosts. Host stars of sub-Saturns with significant eccentricity detected in our analysis are represented as large diamonds. Over plotted is the Terminal Age Main Sequence line. }
\label{fig:hrdiagram}
\end{figure}

We use the derived isochrone mass and radius for calculation of the ``true'' stellar density, $\rho_{\star}$. We report the derived stellar radius and density in Table \ref{tab:final_table}, and show the locations of the host stars on a HR diagram in Figure \ref{fig:hrdiagram}. The majority of our planet-host stars are on the main sequence.  

\subsection{Light Curve Modelling}
\label{ssec:planetmodelling}
We determine the eccentricity of each planet using \tess{} light curves through the ``photoeccentric effect'' \citep{Dawson:2012a}. This involves comparing the ``true'' stellar density--the density derived from our independent stellar modelling as described in Section \S\ref{ssec:stellarmodelling}--labelled $\rho_{\star}$, with the stellar density derived from light curve modelling, $\rho_{\mathrm{circ}}$. We describe this value as $\rho_{\mathrm{circ}}$ as this density is implied assuming the planet follows a circular orbit. The ratio of $\rho_{\star}$ to $\rho_{\mathrm{circ}}$ is related by
\begin{equation}
\centering
    \frac{\rho_{\star}(e,\omega)}{\rho_{\mathrm{circ}}(e, \omega)} = g(e,\omega)^{-3}
\label{eq:photoecc}
\end{equation}
where $g$($e, \omega$) = ($1 + e\mathrm{sin}\omega)/(\sqrt{1-e^2})$ relates the eccentricity, $e$ and the argument of periastron, $\omega$, to the stellar densities (for a full description, see \citealt{Dawson:2012a}). Here, we model the best-fit planet and stellar parameters for inference of the population eccentricity distribution. We exclude 11 candidates with unresolved period aliases, as interpretation of $\rho_{\mathrm{circ}}$ is compromised if the period is incorrect or unknown.\footnote{This notably ruled out TOI-2134c, the first confirmed eccentric sub-Saturn in a multi-planet system, which was first detected as a single-transit event by \tess{} \citep{toi2134}.} 

We model each system using the Hamiltonian Monte Carlo (HMC) algorithm with the $\mathtt{PyMC}$ package \citep{pymc2023}. We use the $\mathtt{exoplanet}$ package to model a quadratic limb-darkening transit model with $\mathtt{KeplerianOrbit}$ \citep{Mandel&Agol:2002, exoplanet:kipping13, exoplanet:luger18}. We use the \tess{} light curves described in Section \S\ref{sec:data} and simultaneously detrend using a Matern-3/2 Gaussian Process (GP) kernel with $\mathtt{celerite2}$ \citep{celerite2}. 
We include only data corresponding to five times the observed transit duration before and after the detected transit center for computational efficiency. 
The Matern-3/2 kernel is described by the function
 \begin{equation}
 \centering
    k(\tau) = \sigma^2 \left(1 + \frac{\sqrt{3}\tau}{\rho}\right) \mathrm{exp}\left(-\frac{\sqrt{3}\tau}{\rho}\right)
 \end{equation}
where $\sigma$ is the amplitude of variability and $\rho$ is the GP timescale. 
Each system's free parameters include the orbital period of the planet (P), the time of inferior conjunction ($\mathrm{T_c}$), the ratio of the radius of the planet to the star ($R_p/R_{\star}$), the planet impact parameter (b), the limb-darkening coefficients described in \citet{exoplanet:kipping13} ($q_1$ \& $q_2$), and the mean stellar density assuming a circular orbit ($\rho_{\mathrm{circ}}$). 

{For $P$ and $\mathrm{T_c}$, we adopt uniform priors centered on the TOI catalog values \citep{TOI}. The prior width spans $\pm10\sigma$ from the catalog uncertainties, with minimum thresholds of $10^{-5}$ days for $P$ and $10^{-2}$ days for $\mathrm{T_c}$ to prevent under-estimated uncertainties.} We impose a log-uniform prior from 
$5.5\times10^{-3}$ to $0.7$
for $R_p/R_{\star}$. The prior on b is bounded uniformly from $-2$ to 2, where we take the absolute values of the samples for our final result. The priors on the parametrized limb darkening parameters $q_1$ and $q_2$ are uniform from 0 to 1. 
We use a log-uniform prior bounded from $10^{-3}$ g $\mathrm{cm}^{-3}$ to $10^{3}$ g $\mathrm{cm}^{-3}$ for $\rho_{\mathrm{circ}}$ {to prevent implicit biases on the impact parameter \citep{Gilbert:2022}}. 

Some planetary candidates in our sample exhibit Transit Timing Variations (TTVs) due to interactions with additional companions in their systems. We manually vet and identified 5 systems showing potential TTVs and fit for their individual transit times separately using \textsc{exoplanet}'s \textsc{TTVOrbit}. We present the result from our TTV modelling in Figure \ref{fig:ttvplots}, where the linear ephemerides are derived from a least-squares model using all the best-fit transit centres. Three systems exhibit significant TTVs: TOI-1136 d \citep{Dai:2023}, HD 28109 c (TOI-282.01) \citep{Drasfield:2022}, and TOI 6109.01. 

For each planet, we ran $\mathtt{PyMC}$'s HMC No-U-Turn Sampler \citep{Hoffman:2011} with four chains each consisting of 5000 tuning steps and a subsequent {5000} samples. The target acceptance is set to 0.99 to reduce divergences occurring due to the complex parameter space between b and $\rho_{\mathrm{circ}}$. We assess the convergence of chains using the Gelman-Rubin diagnostic \citep{Gelman:1992} ($\hat{\mathcal{R}}$), where a value less than 1.01 satisfies the criterion indicating that all chains have converged. We also use a suite of diagnostic statistics including trace plots and corner plots to assess convergence.

\begin{figure*}
    \centering
    \includegraphics[width=0.8\columnwidth]{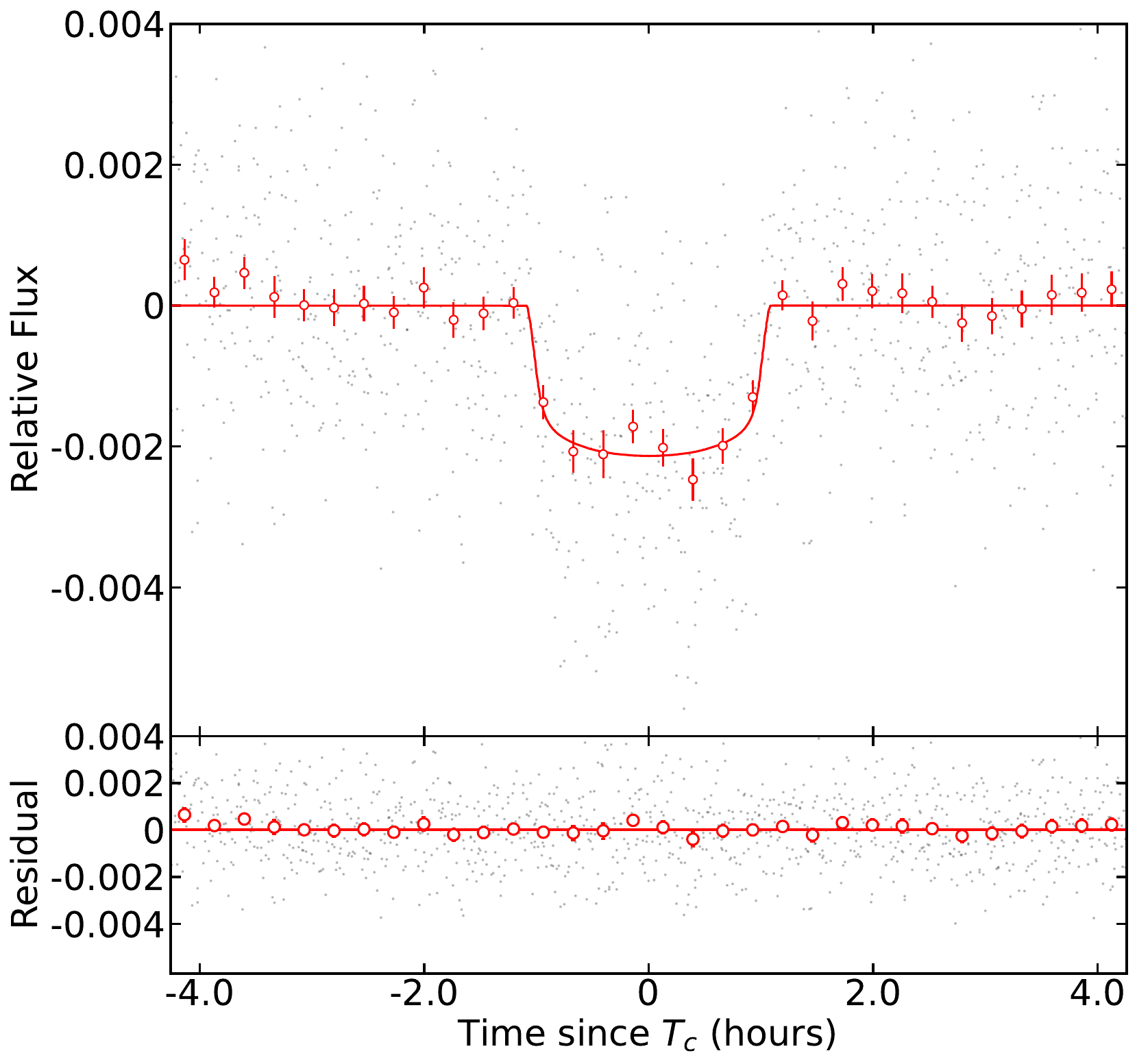}
    \includegraphics[width=0.8\columnwidth]{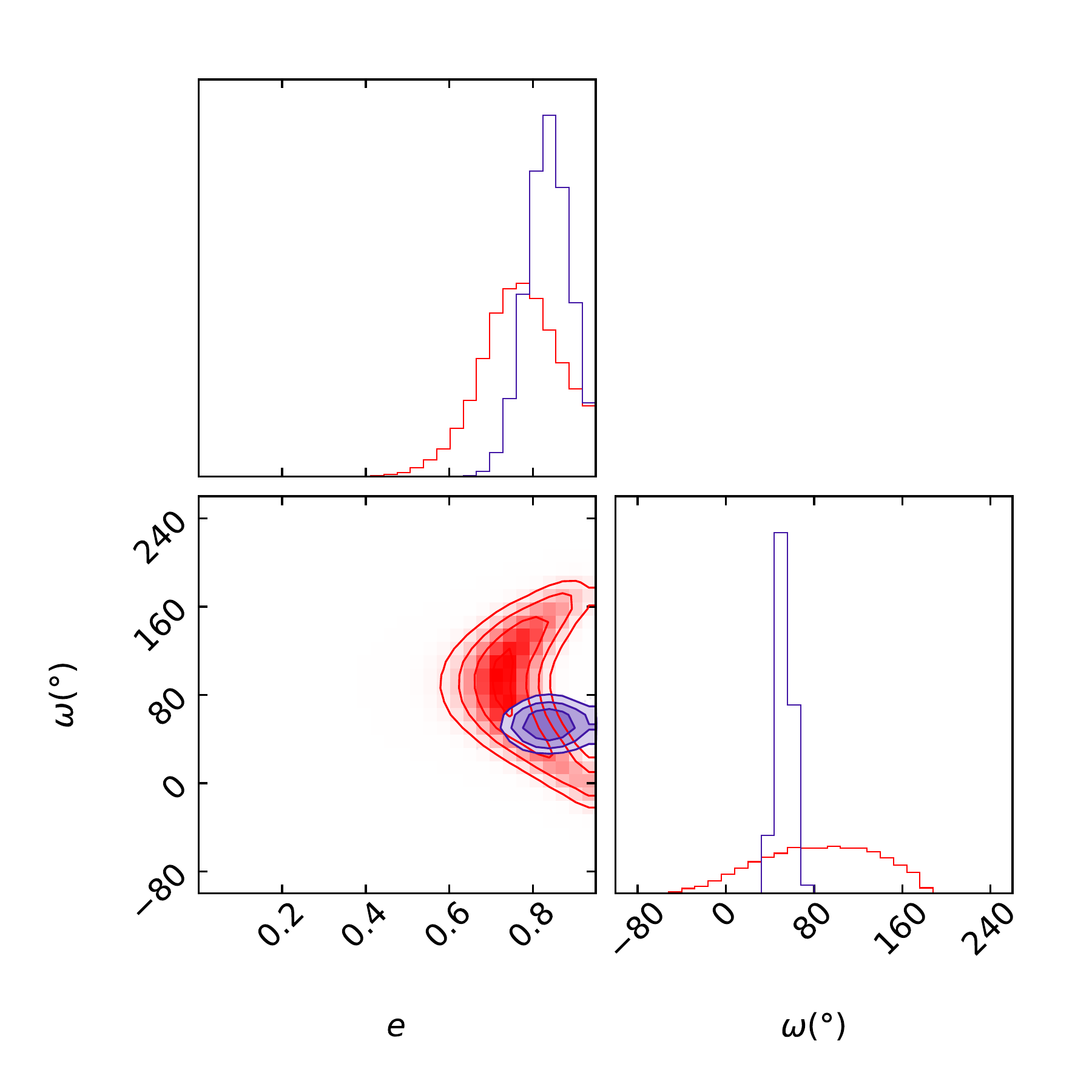}
    \caption{{Left:} Phase-folded light curve and associated residuals of Kepler-1656b with the best-fit planetary transit model overlaid. red points indicate binned data. {Right:} The \tess{} transit photometry derived "photoeccentric effect" posterior distribution for $e$ and $\omega$ with uniform priors (red), with the overlaid radial velocity derived eccentricity solution in blue.}
    \label{fig:k1656}
\end{figure*}

We use our obtained best-fit system parameters to further refine our sub-Saturn sample. Planets exhibiting any of the following characteristics are excluded from the final sample. 
\begin{itemize}
    \item Impact parameter (b) $\ge$ 0.9, due to increased radius uncertainty and higher false positive rate and also compromise the assumptions made for the photoeccentric effect. This removed 15 candidates. 
    \item Best-fit planet radius outside of the defined 4-8 R$_\oplus$ range. These planets, however, are used in robustness testing as described in Section \S\ref{ssec:bootstrap}. An additional {25} candidates were removed after this cut.   
\end{itemize}

{The full sample selection process can be seen in Table \ref{tab:sample_selection}}. Our final sample of warm sub-Saturns consists of {60} single transiting planets and 16 multi-transiting planets included in analysis of the population eccentricity distribution. Their best-fit parameters are presented in Table \ref{tab:final_table}. Individual best-fit transit models compared to each target's light curve are also shown in Figure \ref{fig:lightcurves}.

{Each individual planet now has the required priors--namely $\rho_{circ}$, $\rho_{\star}$ and $P$--to derive their eccentricities, leveraging the ''photoeccentric" effect. We use Equation \ref{eq:photoecc} to infer the $e$-$\omega$ joint posterior for each system using our derived parameters. We fit for $\sqrt{e}{\mathrm{sin}}\omega$ and $\sqrt{e}{\mathrm{cos}}\omega$ in our analysis, placing uniform priors from -1 to 1. We use the $\rho_{\star}$ posterior, along with our fixed best-fit median value of $P$, to correct for the geometric bias for eccentric transiting planets, as described in Section \S\ref{sec:distribution}. The model is initialized using the same settings as earlier described for light-curve fitting. Given the potential asymmetry in posterior distributions—arising from the eccentricity and bimodal nature of the argument of periastron—we present the eccentricity modes and their corresponding 68\% Highest Density Intervals (HDI) for each target in Table \ref{tab:final_table}, rather than the median and 68\% percentiles.}

We find that our derived best-fit system parameters are consistent with previously published warm sub-Saturns. To demonstrate the effectiveness of $\rho_{\mathrm{circ}}$ on constraining eccentricity from \tess{} light curves, we derived $e$ and $\omega$ for Kepler-1656b \citep{Brady:2018}, a confirmed high-eccentricity warm sub-Saturn ($e=0.838_{-0.029}^{+0.045}$) using a uninformative uniform prior on both parameters. Our fit yields a best fit eccentricity of ${0.79_{-0.12}^{+0.08}}$, within 1$\sigma$ of the published value derived from radial velocities. The best fit photoeccentric model and phase-folded Kepler-1656 \tess{} light curve, as well as the posteriors of $e$--$\omega$ for the planet, is shown in Figure \ref{fig:k1656}. 

\section{The Eccentricity Distribution}
\label{sec:distribution}
\subsection{Hierarchical Bayesian modelling of the eccentricity distribution}

Due to the unknown nature of sub-Saturn formation, we fit a range of model distributions to capture both single and multiple formation pathways. We conduct a Hierarchical Bayesian Modelling (HBM) inference using three populations: single-transiting, multi-transiting and the combined total population. 
We fit each model as a Beta distribution as a non-physically motivated distribution with two-parameter flexibility and bounding to the domain of 0 to 1 \citep{Hogg:2010, Kipping:2013}. We also extend the two-component Mixture Gaussian distribution from \citet{Dong:2021} by introducing the Mixture Beta model. Given the Beta distribution’s flexibility of allowing for asymmetric uncertainties around a mean, it serves as a useful tool in representing potentially asymmetric two-component distributions. Formally it can be written as

\begin{equation}
    p\left(e \,\big|\, w_1, w_2, \alpha_1, \alpha_2, \beta_1, \beta_2\right) = w_1\,\mathcal{B}\left(e \,\big|\, \alpha_1, \beta_1\right) + w_2\,\mathcal{B}\left(e \,\big|\, \alpha_2, \beta_2\right)
\end{equation}
where \( w_1 \) and \( w_2 \) represent the fractions for each mode of the distribution and sum to one, and where \( \mathcal{B}(e|\alpha, \beta) \) indicates a Beta distribution with hyperparameters \( \alpha \) and \( \beta \).

We construct our hierarchical model following the framework in \citet{Dong:2021}, but we reparameterize the priors in the Beta models, using the mean (\( \mu \)) and concentration (\( \kappa \)) as reparametrizations of the Beta distribution hyperparameters as in \cite{Dong:2023}. For each Beta component, we have
\begin{equation}
    p\left(e_i \,\big|\, \mu_e, \kappa_e\right) \sim \mathcal{B}\left(e_i \,\big|\, \mu_e, \kappa_e\right)
\end{equation}
\begin{equation}
    \mu_e \sim \mathcal{U}\left(0, 1\right)
\end{equation}

\begin{equation}
    \mathrm{log}(\kappa_e) \sim \mathcal{N}\left(3, 1\right)
\end{equation}
where \( \mathcal{U} \) denotes a uniform distribution and \( \mathcal{N} \) represents a normal distribution. We also apply a transit probability weighting factor to correct for the observation bias regarding the eccentricities of transiting planets \citep{Burke_2008, Winn:2010, Kipping_2014b}. This is described as
\begin{equation}
    p\left(\text{obs}_i \,\big|\, \rho_{\star, i},\, e_i,\, \omega_i,\, P_i\right) =
    \begin{cases} 
        \dfrac{1}{\rho_{\star, i}^{1/3} P_i^{2/3}} \cdot \dfrac{1 + e_i \sin \omega_i}{1 - e_i^2} & \text{if } e_i < e_{\text{max},i}, \\ 
        0 & \text{if } e_i \geq e_{\text{max},i}.
    \end{cases}
\end{equation}
where $e_{\text{max},i} = 1 - R_{\star, i}/a_i$ is the maximum eccentricity a planet can attain {before colliding with its host star}. 

\begin{figure*}
    \centering
    \includegraphics[width=\textwidth]{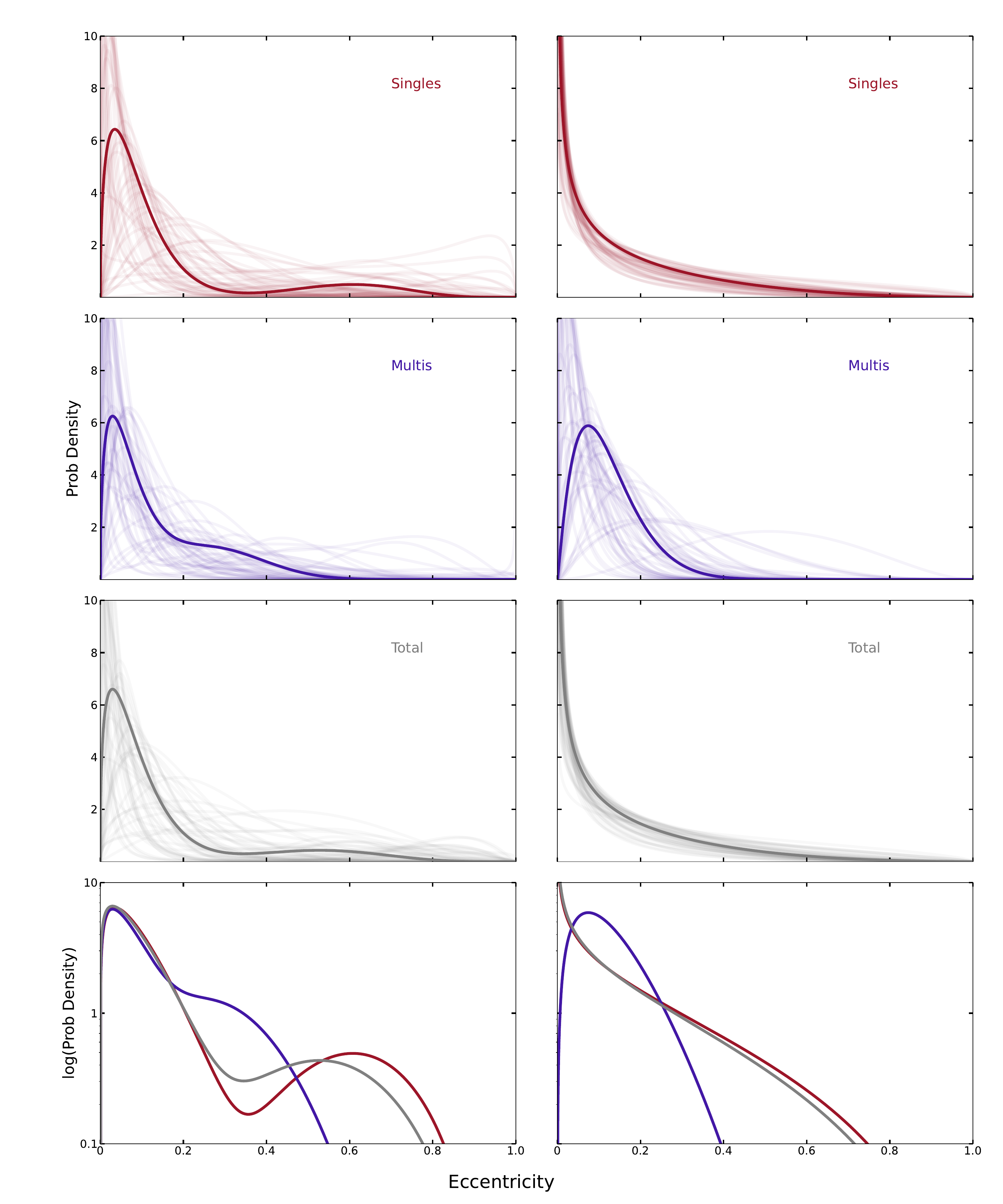}
    \caption{Comparison of inferred eccentricity distributions for the single-transiting (red) and multi-transiting (dark purple) warm sub-Saturn population with both the Beta Mixture (Left panel) and Beta (Right panel) functional forms, along with 50 random draws from the 1$\sigma$ uncertainties of the population parameters. For references, the eccentricity distributions from the total sample is plotted in grey. The bottom most panel over-plots the best-fit distribution from all three samples in log y-axis scale to show the differences.}
    \label{fig:edist}
\end{figure*}

\begin{figure}
    \centering
    \includegraphics[width=0.99\linewidth]{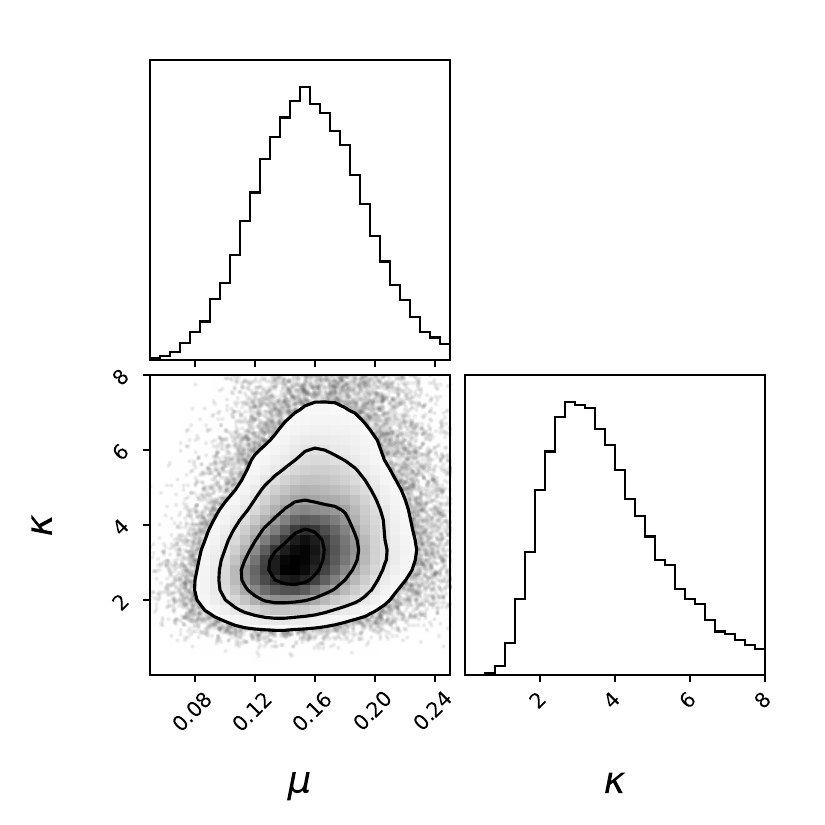}
    \caption{Corner plot showing the posteriors of the fitted hyperparameters ($\mu$, $\kappa$) total sample's eccentricity distribution assuming a Beta Distribution.}
    \label{fig:beta-corner}
\end{figure}

\begin{figure*}
    \centering
    \includegraphics[width=0.99\linewidth]{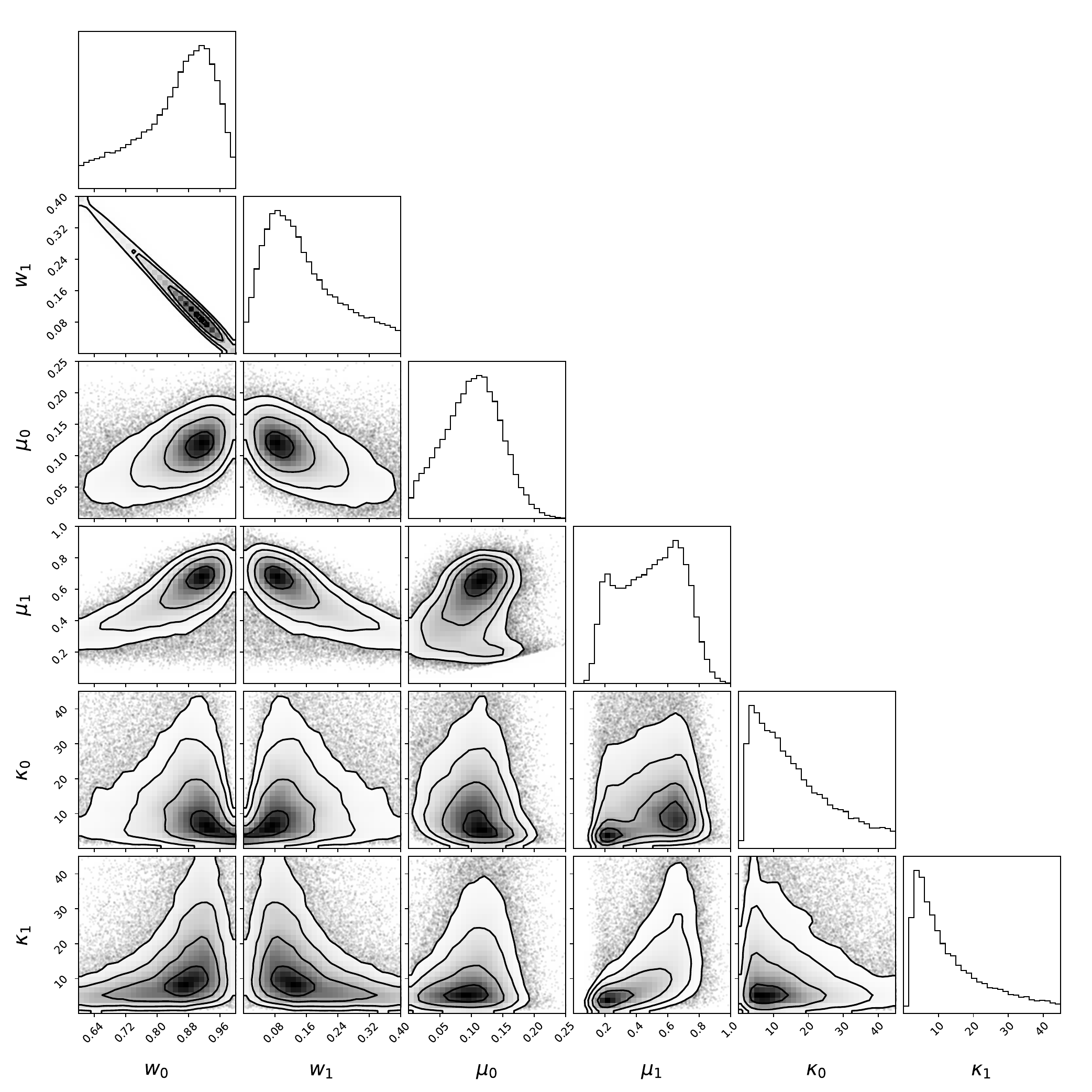}
    \caption{Corner plot showing the posteriors of the eccentricity distribution hyperparameters for the total sample assuming a Beta-Mixture model.}
    \label{fig:betaMixture-corner}
\end{figure*}

The hierarchical model is built using $\mathtt{PyMC}$ \citep{pymc2023}. To reduce computational time, we assume the best-fit MCMC samples for each planetary system are drawn from a Gaussian distribution and assume the orbital period to be fixed at the best-fitted value from Section \S\ref{ssec:planetmodelling}, given the typical period uncertainties cause a negligible difference to the transit probability weighting. We set a uniform target acceptance of 0.99 and sample from four chains, each with 40,000 tuning steps and 10,000 draws. The three resulting modelled distributions are shown in Figure \ref{fig:edist}, with the corner plots of the total sample's Beta and Beta Mixtures fitted parameters shown in {Figures} \ref{fig:beta-corner} and \ref{fig:betaMixture-corner}. {The hyperparameters derived from our analysis are reported in Table \ref{tab:hbm_parameters_table}.}

While we do not formally compute the Bayesian evidences of each models in this work, we validated the robustness of both the single and multi-component Beta distributions on capturing the eccentricity distribution of the warm sub-Saturn population using both Leave-One-Out Cross-validation (LOO, \citealt{Vehtari:2015}) and the Widely-Applicable Information Criterion (WAIC, \citealt{Watanabe2012AWA}). LOO cross-validation involves dividing the observed data into training and holdout sets, and iteratively fitting the model to the data, evaluating the goodness of fit with the holdout data. WAIC uses the log pointwise posterior predictive density to estimate the out-of-sample expectation, correcting for the effective number of parameters to account for model overfitting. 

We find a $\Delta$LOO of {0.88} ($\Delta$ WAIC of {0.32}) between the Beta Mixture model and the Beta model, {the statistical significance of which is ${0.7\sigma}$ (${0.6\sigma}$), indicating that both models perform similarly in explaining the distribution.} We also note that there is a strong covariance between $\kappa_0$ and $\mu_1$ in Figure \ref{fig:betaMixture-corner}, which may be an indication of a preference for the single-component model rather than the Mixture. {We discuss the formation of high-eccentricity (${e > 0.3}$) sub-Saturns as a potential alternate origin channel in Section \S\ref{sec:discussion}, without asserting that these systems are outliers or a distinct population. }


\begin{table*}
\centering
\fontsize{6}{8}\selectfont
\caption{Best fit parameters from our HBM model for the Beta and Beta Mixture distributions derived from our planet samples. We report the median and 68\% Highest Density Interval for each hyperparameter. }
\label{tab:hbm_parameters_table}
\setlength{\extrarowheight}{2pt}
\renewcommand*{\arraystretch}{1.5}
\begin{tabular}{llllllll}
\hline\hline
Distribution & & Total Sample & Total (bootstrap) & Single Only & Single (bootstrap) & Multi Only & Multi (bootstrap) \\
\hline\hline
\multirow{6}{*}{Beta}
 & Fitted:\\
 & $\mu$ & $0.156_{-0.034}^{+0.037}$ & $0.153_{-0.032}^{+0.049}$ & $0.169_{-0.042}^{+0.040}$ & $0.144_{-0.040}^{+0.031}$ & $0.131_{-0.072}^{+0.057}$ & $0.130_{-0.093}^{+0.057}$ \\
 & $\kappa$ & $3.6_{-1.8}^{+1.2}$ & $3.7_{-1.7}^{+1.0}$ & $3.5_{-1.8}^{+1.2}$ & $4.2_{-1.8}^{+1.8}$ & $17_{-15}^{+11}$ & $21_{-15}^{+12}$ \\
 & Derived:\\
 & $\alpha$ & $0.56_{-0.34}^{+0.22}$ & $0.57_{-0.27}^{+0.29}$ & $0.58_{-0.37}^{+0.23}$ & $0.67_{-0.35}^{+0.23}$ & $2.1_{-2.1}^{+1.5}$ & $2.0_{-1.6}^{+1.6}$ \\
 & $\beta$ & $3.1_{-1.5}^{+1.0}$ & $3.0_{-1.0}^{+0.9}$ & $2.9_{-1.5}^{+1.0}$ & $3.7_{-1.6}^{+1.5}$ & $15_{-13}^{+9}$ & $19_{-14}^{+8}$ \\
\multirow{12}{*}{Beta Mixture}
 & Fitted:\\
 & $w_1$ & $0.84_{-0.10}^{+0.15}$ & $0.83_{-0.11}^{+0.16}$ & $0.84_{-0.09}^{+0.13}$ & $0.84_{-0.10}^{+0.15}$ & $0.71_{-0.17}^{+0.29}$ & $0.73_{-0.17}^{+0.27}$ \\
 & $\mu_1$ & $0.103_{-0.045}^{+0.047}$ & $0.093_{-0.052}^{+0.053}$ & $0.103_{-0.046}^{+0.050}$ & $0.094_{-0.060}^{+0.050}$ & $0.087_{-0.085}^{+0.035}$ & $0.082_{-0.082}^{+0.043}$ \\
 & $\kappa_1$ & $16_{-15}^{+10}$ & $15_{-14}^{+10}$ & $17_{-16}^{+10}$ & $16_{-14}^{+10}$ & $19_{-18}^{+12}$ & $19_{-17}^{+12}$ \\
 & $w_2$ & $0.16_{-0.15}^{+0.10}$ & $0.17_{-0.16}^{+0.11}$ & $0.16_{-0.13}^{+0.09}$ & $0.16_{-0.15}^{+0.10}$ & $0.29_{-0.29}^{+0.17}$ & $0.27_{-0.27}^{+0.17}$ \\
 & $\mu_2$ & $0.51_{-0.19}^{+0.25}$ & $0.47_{-0.30}^{+0.16}$ & $0.58_{-0.16}^{+0.23}$ & $0.53_{-0.26}^{+0.25}$ & $0.28_{-0.18}^{+0.13}$ & $0.28_{-0.20}^{+0.14}$ \\
 & $\kappa_2$ & $13_{-11}^{+10}$ & $12_{-11}^{+9}$ & $15_{-13}^{+11}$ & $12_{-11}^{+9}$ & $21_{-18}^{+13}$ & $20_{-18}^{+13}$ \\
 & Derived:\\
 & $\alpha_1$ & $1.4_{-1.3}^{+1.0}$ & $1.1_{-1.1}^{+0.9}$ & $1.5_{-1.4}^{+1.1}$ & $1.3_{-1.3}^{+1.0}$ & $1.5_{-1.5}^{+1.2}$ & $1.4_{-1.5}^{+1.2}$ \\
 & $\beta_1$ & $14_{-13}^{+9}$ & $13_{-12}^{+9}$ & $15_{-14}^{+9}$ & $14_{-14}^{+9}$ & $18_{-16}^{+11}$ & $17_{-15}^{+11}$ \\
 & $\alpha_2$ & $6.3_{-6.1}^{+5.9}$ & $5.6_{-5.5}^{+5.3}$ & $8.3_{-8.1}^{+7.2}$ & $6.1_{-6.0}^{+6.0}$ & $5.8_{-5.6}^{+4.3}$ & $5.7_{-5.5}^{+4.4}$ \\
 & $\beta_2$ & $5.7_{-4.5}^{+3.7}$ & $5.8_{-4.9}^{+3.6}$ & $5.7_{-4.8}^{+3.5}$ & $5.3_{-4.7}^{+4.0}$ & $13_{-13}^{+8}$ & $12_{-12}^{+8}$ \\
\hline
\end{tabular}
\end{table*}

\subsection{Robustness Check of the Distributions with Bootstrapping}
\label{ssec:bootstrap}
To test how astrophysical false positives (FP) or planets possibly not being sub-Saturn-sized affect our derived eccentricity distribution, we conducted a robustness using bootstrapping analysis. 

We estimate the FP rate of all unconfirmed/unpublished planets in our sample using \tess{} Follow-up Observing Program \citep{TFOP} Sub-Group 1 and 2 (SG1 and SG2) statistics. SG1 \citep{Collins:2019} uses ground-based seeing limited photometry to rule out astrophysical false positive scenarios due to nearby and background eclipsing binaries (NEB and BEB), while SG2 \citep{Quinn:2019} use reconnaissance spectroscopy to rule out false positive scenarios due to on-target binaries and some hierarchical triples. 

Among all 993 historical TOIs with a radius between 4-8 $R_{\oplus}$, SG1 and SG2 followed up and identified 368 systems to be false positives.  
We assign the false positive rate of the planetary candidates based on the amount of photometric follow-up effort they already obtained. For planetary candidates that did not receive sufficient photometric follow-up (38 systems, labelled as PC), we assume a false positive rate of 40\%, which is $\sim5$\% higher than the false positive rate of sub-Saturns derived from TFOP statistics. This is likely an overestimation of the false positive rate, given the statistics are dominated by planets on orbital periods less than 8 days. Due to the strong dependence of background eclipsing binary occurrence rate on their orbital period \citep{Fressin:2013}, shorter period planets typically have a higher false positive rates compared to longer period planets. For planetary candidates with seeing-limited photometric follow-up and with all nearby stars cleared of NEB scenarios, we use the historic false positive rate (5\%, 34 systems, labelled as Cleared Candidate (CC)) of TOIs that become false positives after additional follow-up data from spectroscopy and adaptive optics are obtained. 

We use the median radius of each planet as well as the lower and upper 1$\sigma$ uncertainties derived from our star and planet modelling to estimate the probability of the planet falling within the 4-8 $R_{\oplus}$ range. 
We also use the impact parameter's median and 1$\sigma$ asymmetric uncertainties to assign a probability of each target being b$<0.9$. 

We include candidates that were removed due to having radii outside the 4-8 $R_{\oplus}$ range in our bootstrapping analysis. We do not, however, use targets with best-fit impact parameters exceeding 0.9 as their implied stellar densities and radii are heavily degenerate.

The final probability of a planet drawn in the bootstrapping is the combination with the above probabilities assuming they are independent. We resample the planets 10 times and combine the posterior distribution of the hyperparameters to obtain the final eccentricity distribution from bootstrap. The best-fit parameters for the hyper-parameters of each distribution from the bootstrap is shown in Table \ref{tab:hbm_parameters_table} and consistent with our findings.

\section{Results and Discussion}
\label{sec:discussion}
\subsection{Implications of the Eccentricity Distribution}

\begin{figure*}
    \centering
    \includegraphics[width=\textwidth]{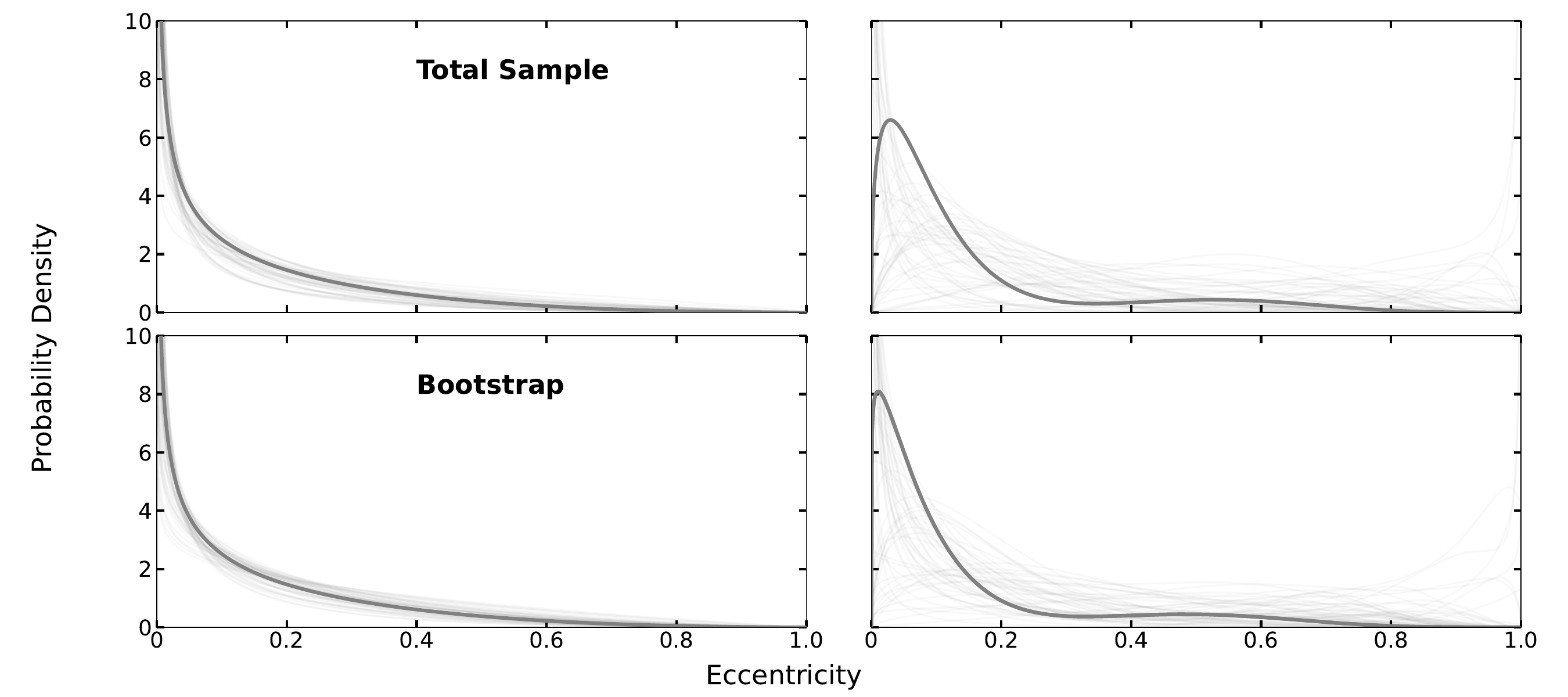}
    \caption{The two functional forms of the inferred eccentricity distribution for the warm sub-Saturn population, along with 50 random draws from the 1$\sigma$ uncertainties of the population parameters. {Left:} The single-component Beta distribution. {Right:} Two-component Beta Mixture distribution. The top panel is the default result with our total sub-Saturn samples. The bottom panel shows the result from our bootstrap experiment for comparison.}
    \label{fig:bootstrap}
\end{figure*}

\begin{figure*}
\includegraphics[width=0.99\textwidth]{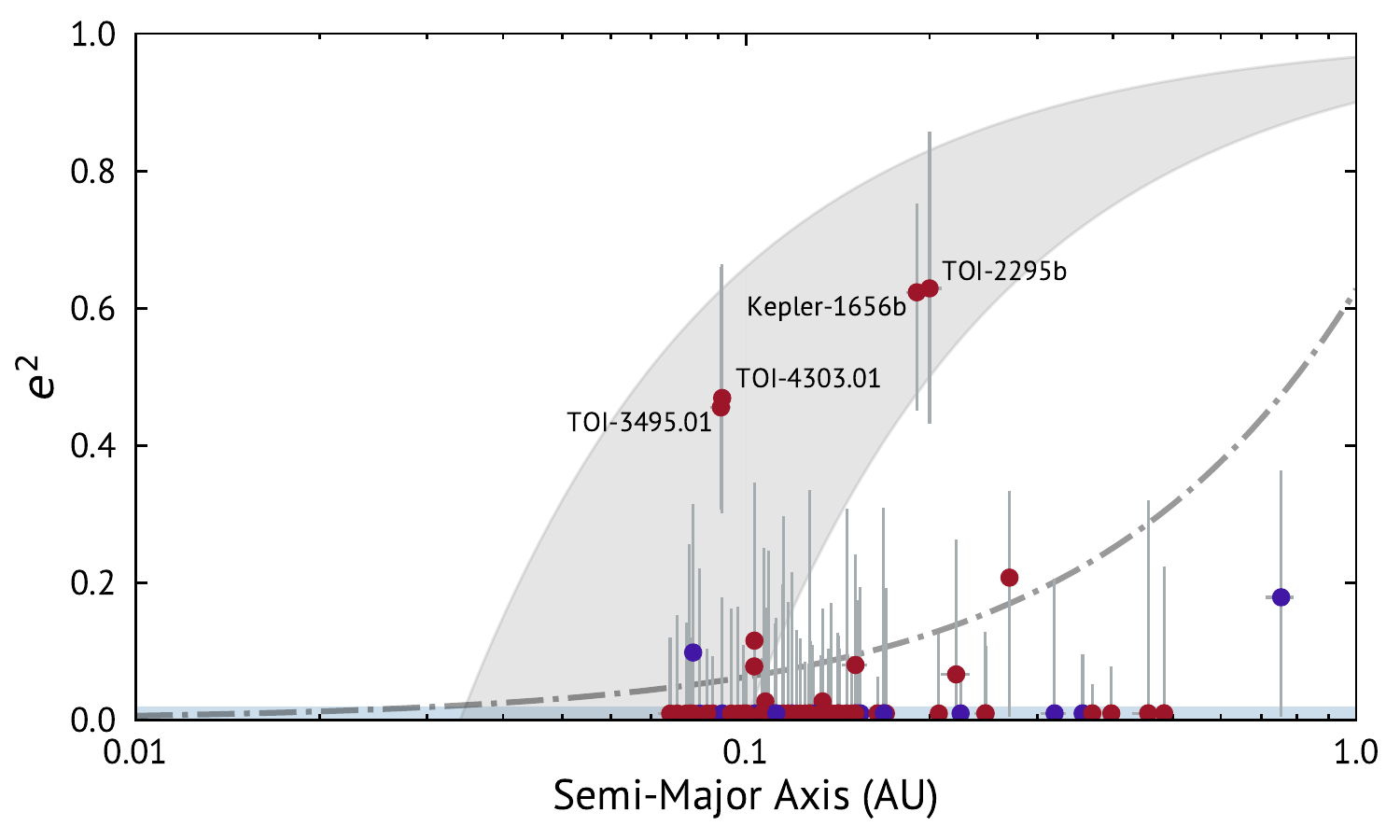}
\caption{$e^2$ vs. semi-major axis for the total sample. The eccentricity is scaled to $e^2$ for visualization purposes. Eccentricities plotted for individual sub-Saturns are derived using a uniform prior on both eccentricity and argument of periastron. Single-transiting sub-Saturns are plotted in red, while multi-transiting sub-Saturns are plotted in dark purple. The shaded region represents the parameter space occupied by planets that likely evolved through high-eccentricity migration. The boundaries are determined following angular momentum tracks as in \citet{toi3362}. The dash-dotted line represents the maximum eccentricity a planet can be excited to from planet-planet scattering. Both the high-eccentricity migration boundaries and the planet-planet scattering line are estimated assuming planet mass of 0.1 $M_J$ and radius of 0.6 $R_J$ around a 1 solar mass host star following Eq 10 from \citet{Dawson:2018}.}\label{fig:ecc_vs_a}
\end{figure*}

Our HBM analysis, as seen in Figure \ref{fig:edist}, reveals that both a Beta distribution and a Mixture Beta model describe the warm sub-Saturn eccentricity distribution comparably well. Using the method described in Section \S\ref{ssec:bootstrap}, we also obtained the bootstrapped distribution for each sample. We compare the derived distribution of the total sample to the bootstrapped analysis in Figure \ref{fig:bootstrap}, and report the best-fit values in Table \ref{tab:hbm_parameters_table}. The agreement between our results and the bootstrapping confirms that our derived distribution is robust to our sample selection process.   

The Beta model yields a mean eccentricity of ${0.156_{-0.034}^{+0.037}}$ (median ${e\approx0.09}$) with a variance of $0.03\pm0.01$. This mean eccentricity aligns closely with that of warm Jupiters in \tess{} \citep[$\sim0.19$, ][]{Dong:2021}, while falling between the derived values for Kepler single-transiting (0.26-0.32) and multi-transiting ($<0.07$) systems \citep[e.g.][]{Xie:2016, VanEylen:2019}.

Our best-fit two-component Beta Mixture distribution inference suggests that the majority (${\sim84\%}$) of warm sub-Saturns have low eccentricities (${\bar{e}=0.103_{-0.045}^{+0.047}}$), while a small but significant fraction (${\sim16\%}$) exhibit high eccentricities with mean ${0.51_{-0.19}^{+0.25}}$. Though this high-eccentricity mode exceeds the average eccentricity of Kepler single-transiting super-Earths and sub-Neptunes, both populations remain consistent within 1$\sigma$. \citet{VanEylen:2019} modelled a Rayleigh/half-Gaussian mixture model for small planets (R$< 6 R_{\oplus}$, P > 5 days) and also identified a moderate eccentricity component ($\bar{e}\sim0.33$). The high eccentricity mode in our sample includes a handful of planets/candidates such as Kepler-1656b, with eccentricities comparable to highly eccentric warm Jupiters such as TOI-3362 b \citep[$e=0.720_{-0.016}^{+0.016},$][]{toi3362_new}. These planet/candidates will be discussed individually in \S \ref{sec:candidates}. These extreme eccentricities are consistent with the parameter space sculpted by high eccentricity tidal migration, and are difficult to explain through typical planet-planet interactions.

The Beta mixture analysis of single- versus multi-transiting systems reveals similar low-eccentricity modes, but distinct high-eccentricity behavior. Single-transiting systems exhibit a significantly higher mean eccentricity (${0.58_{-0.16}^{+0.23}}$) in their high-eccentricity mode, while multi-transiting systems favor a more modest eccentric mode (${0.28_{-0.18}^{+0.13}}$). Given the small number statistics in the multi-transiting planets sample, it is unclear if this modest eccentricity mode is real. In contrast, the high eccentricity mode in the single-transiting distribution seems to be the major contributor to the entire high eccentricity sample.

We use CORBITS \citep{CORBITS} to test the possibility that our low-eccentricity single transiting sub-Saturns are mostly sub-Saturns in multi-planet systems. CORBITS is used to estimate the geometric transit probability of a sub-set of planets in a multi-planet system. We take the orbital parameters of our sub-Saturns in multi-transiting systems \footnote{only systems with two transiting planets are used, given the uncertainty in the mutual inclination distribution for higher multiplicity systems. We also omitted systems with outer warm Jupiter companions.}, assuming these systems have similar mutual inclination distributions compared to the typical Kepler multi-planets \citep{Fabrycky:2014}. With random phase viewing angles taken into account, we estimate that to produce the multi-transiting sub-Saturn systems we have in our sample, there should be $\approx 120$ single-transiting sub-Saturn systems observed by \tess{} from the same distribution. Given we only have {100} single-transiting sub-Saturns in our initial sample, we conclude it is most likely that the majority of single-transiting sub-Saturns are from the same population as the multi-transiting sub-Saturns. 

In summary, our analysis of the \tess{} warm sub-Saturn population reveals that multiple eccentricity excitation mechanisms likely sculpt these planets. A small but significant fraction of sub-Saturns exhibit high eccentricities, consistent with high-eccentricity migration pathways, which is similarly found in the warm Jupiter population. 
A uniform analysis comparing sub-Saturns with smaller planets would be useful to confirm any eccentricity-radius trend. Additionally, highly eccentric sub-Saturns remain understudied; future discoveries and detailed characterization of these systems will be vital for constraining their formation conditions.

\subsection{Possible Biases and Robustness of the Inferred Distributions}

{Our derived eccentricity distribution is corrected for the transit probability. However, to obtain the intrinsic eccentricity distribution of the planet population, future work needs to also correct for the detection completeness of the planets, especially those due to the limited baseline and complex window functions of \tess. The detection incompleteness has the strongest impact on long-period planets, and thus ignoring the incompleteness may lead to underestimating the fraction of planets on eccentric orbits. Most of the eccentricity excitation mechanisms predict larger eccentricities for longer-period planets \citep{Chatterjee:2008}. }

{A small subset of planets were affected by detection incompleteness, as they lack a unique period solution when analyzing only \tess{} data. In total, eight candidates and three known planets from the TOI list are affected. The eccentricity of these planets cannot be constrained from the photoeccentric effect without a confirmed orbital period. However, we do have period and eccentricity information for the three known planets--TOI-5678b, Kepler-396c and TOI-2134c--from previous publications \citep{TOI5678, KEPLER396, KEPLER396_2, toi2134}. Here, we examine the eccentricity distributions with these three planets included and compare them with our fiducial eccentricity distribution. The final sample yields a mean eccentricity hyperparameter of $0.166_{-0.036}^{+0.034}$ for the Beta model. Performing this analysis with the Beta Mixture also resulted in $1\sigma$ agreement.} 

\subsection{Highlight of Individual Systems}
\label{sec:candidates}
In analysis of the individual target eccentricities, we have identified {four} systems with their orbital eccentricities significantly excluding circular orbits ($>4\sigma$ detection). Each system will be discussed briefly. 

{{Kepler-1656b (TOI-4584.01):}}
We find a best-fit planet radius and period of ${4.82_{-0.27}^{+0.26} R_{\oplus}}$ and ${\approx31.58}$ days, respectively, for Kepler-1656b. The host star, Kepler-1656, has a derived stellar radius and mass of ${R_{\star}=0.993\pm0.027 R_{\odot}}$ and ${M_{\star}=0.93\pm0.11 M_{\odot}}$, respectively. We derive an eccentricity of ${e=0.79_{-0.12}^{+0.08}}$, within $1\sigma$ agreement of its previously confirmed RV eccentricity of $e=0.838_{-0.029}^{+0.045}$ \citep{angelo:2022}. The planet has a measured mass of $M_p=47.8_{-3.3}^{+6.2} M_{\oplus}$, making it a strong candidate for a merger product \citep{Brady:2018, Millholland:2020}.  The orbital properties of Kepler-1656b place the planet firmly in the HEM regime (seen in Figure \ref{fig:ecc_vs_a}), indicating it may be in the process of tidal circularization. The presence of an eccentric ($e\sim0.5$, $M_p\sim120 M_{\oplus}$) non-transiting Jovian companion on a long-period orbit \citep{angelo:2022} supports migration through HEM. {Recently, the planet's sky-projected orbital obliquity was constrained to $35.0^{+14.9}_{-21.6}$ deg ($<50$ deg at 95\% confidence, \citealt{Rubenzahl:2024}). They find the orbital obliquity of the planet to be consistent with evolution through coplanar HEM \citep{Petrovich:2015a}, contrasting the large number of isolated sub-Saturns with high obliquities \citep{Radzom:2024}. However, if the two planets instead exhibit a high mutual inclination, Kepler-1656b may instead be a snapshot of high-eccentricity oscillations \citep{Rubenzahl:2024}. }

{TOI-2295b (TOI-2295.01):}
Our analysis finds {TOI-2295b} to have a radius of ${4.7_{-0.3}^{+2.5} }R_{\oplus}$ and period of ${\approx30.03}$ days. The host star (TIC 48018596) is slightly evolved (${R_{\star}=1.436\pm0.040 R_{\odot}}$, ${M_{\star}=1.19\pm0.17 M_{\odot}}$, ${T_{\rm eff}=5800\pm100}$ K). We infer a significant eccentricity of ${e=0.85_{-0.18}^{+0.09}}$. {During manuscript submission, TOI-2295b was independently confirmed through radial velocity measurements, revealing it to be a grazing, moderately eccentric Jovian planet \citep{Heidari:2024}. While our derived impact parameter agrees with the published value of $b\approx1$ to within $1\sigma$, the available \tess{} photometry prior to sector 70 cannot definitively constrain solutions with $b\ge0.9$. To alleviate concern regarding the contribution of this planet on the population interpretation, we found our derived eccentricity agrees to within 2$\sigma$ of the accepted value of $e={0.334}_{-0.012}^{+0.012}$ \citep{Heidari:2024} with a 2$\sigma$ lower limit of 0.29. This large uncertainty was properly accounted for in our hierarchical Bayesian model through the modelling of $\rho_{circ}$, which in turn depends on $b$. We also performed an independent HBM without TOI-2295b to assess its influence on our results. The Beta model for the total (single) sample yields a mean eccentricity of ${0.153_{-0.037}^{+0.032}}$ (${0.162_{-0.041}^{+0.038}}$). We also find the Mixture to be within 1$\sigma$ for both the final and single only runs. Intriguingly, the TOI-2295 system also hosts a massive, moderately eccentric outer companion (TOI-2295c, $m_csini\approx5.6 M_{J}$, $P_c\approx$ 967 days, $e_c=0.194_{-0.012}^{+0.012}$). While not a sub-Saturn, TOI-2295b may represent an interesting case of high-eccentricity migration (HEM) driven by past dynamical interactions.}

{TOI-3495.01 \& TOI-4303.01:}
{We have identified two candidates with high eccentricities of around 0.6. These two candidates, TOI-3495.01 \& TOI-4303.01, are particularly noteworthy as to date, no transiting planet with a period of $<10$ days and $e\sim0.7$ or higher have been detected. These two candidates both fulfil this quality, offering a unique perspective into the formation mechanisms at play. TOI-3495.01 (${R_p=4.07_{-0.35}^{+0.33} R_{\oplus}}$, ${P\approx8.41}$ days) is found around a star near the TAMS line (${R_{\star} = 1.932\pm0.055 R_{\odot}}$, ${M_{\star} = 1.42\pm0.17 M_{\odot}}$, ${T_{\rm eff} = 6400\pm230}$ K), and has an inferred eccentricity of $0.68_{-0.10}^{+0.16}$. SG1 photometry has ruled out the transit event for all nearby stars using a variable aperture. SOAR HRCam \citep{HRCam} adaptive optics imaging detected no nearby stars within 7.6 magnitudes of the target star beyond 1\arcsec. We derived the eccentricity of TOI-4303.01 (${4.46_{-0.46}^{+0.42} R_{\oplus}}$, ${P\approx8.61}$ days) to be $0.69_{-0.10}^{+0.16}$. The candidate orbits an early-F star (${R_{\star} = 1.768\pm0.050 R_{\odot}}$, ${M_{\star} = 1.37\pm0.14 M_{\odot}}$, ${T_{\rm eff} = 6900\pm130}$ K) and has yet to be cleared by SG1 follow-up.  }

In total, 7 confirmed planets in our sample have additional eccentricity constraints from radial velocity data. Four of the planets have measured eccentricities consistent with a circular orbit, consistent with our low eccentricity solution. Three of the planets have non-zero eccentricities. They are Kepler-1656b, and two other planets with modest eccentricities: TOI-257 b \citep[$e=0.24^{+0.04}_{-0.07}$][]{Addison:2021}, and TOI-421 c, \citep[$e=0.19^{+0.05}_{-0.04}$][]{Krenn:2024}. Photoeccentric analysis is not sensitive enough to identify individual planets with modest eccentricities. Planets such as TOI-257 b and TOI-421 c are currently consistent with the low-eccentricity mode of our Beta Mixture model. Future high-precision eccentricity measurements for more planets in the low-eccentricity mode will help identify if these modest eccentric planets need to be described by additional modes in the Beta Mixture model.  

\section{Summary}
We assembled an initial magnitude-limited sample of warm sub-Saturn candidates, defined as planets between 4 and 8 Earth-radii with periods between 8--200 days, brighter than \tess-band magnitude of 12 using data from the first five years of \tess.  We uniformly derived stellar properties from modelling of the broadband spectral energy distribution for each target (\ref{fig:hrdiagram}). We simultaneously detrended each candidates light curve with a Gaussian Process and fit for the best-fit planetary parameters. We vetted and fitted for transit timing variations in five systems (Figures \ref{fig:lightcurves} and \ref{fig:ttvplots}). The planetary properties and relevant stellar properties are reported in Table \ref{tab:final_table}. We applied a series of cuts to the sample, including removal of high impact parameters, potential period aliases and targets outside the 4--8 $R_{\oplus}$ range, resulting in a final sample of {76} candidates. We then uniformly derived the eccentricity of all targets in our sample (Figure \ref{fig:ecc_vs_a}). We performed Hierarchical Bayesian Modelling using our derived planet and stellar parameters to infer the eccentricity distribution of warm sub-Saturns (Figure \ref{fig:edist}), assuming functional forms of a Beta and Mixture Beta distribution. Each of the single transiting, multi-transiting and combined populations were modelled to account for differing distributions between subpopulations. The final hyperparameters are reported in Table \ref{tab:hbm_parameters_table}. To verify the inferred distributions were robust to false positives, we conducted a bootstrapping analysis (Figure \ref{fig:bootstrap}). We also report the bootstrap hyperparameters in Table \ref{tab:hbm_parameters_table}. The implications of the inferred eccentricity distributions were discussed, where we found that high eccentricity tidal migration is likely present for some warm sub-Saturns, while planet-planet scattering is less constrained (Figure \ref{fig:ecc_vs_a}). We found that the single transiting warm sub-Saturns contribute to most of the high eccentricity mode, while multi-transiting sub-Saturns exhibit a modest mean eccentricity as a population. We also highlighted three systems with significant detections of moderate to high eccentricities. 

\section{Acknowledgements}
\label{acknowledgements}
We respectfully acknowledge the traditional custodians of the lands on which we conducted this research and throughout Australia. We recognize their continued cultural and spiritual connection to the land, waterways, cosmos, and community. We pay our deepest respects to all Elders, present and emerging, and the people of the Giabal, Jarowair, and Kambuwal nations, upon whose lands this research was conducted.

We are grateful to the anonymous referee for their insightful comments which have improved the manuscript. 

T.F. thanks R.W. and B.N. for their helpful comments during the draft manuscript process.

G.Z. thanks the support of the ARC Future program FT230100517.

C.H. thanks the support of the ARC DECRA program DE200101840.

E.N. acknowledges the PhD scholarship provided by the ARC discovery grant DP220100365.

The Flatiron Institute is a division of the Simons foundation.

Funding for the TESS mission is provided by NASA's Science Mission Directorate. KAC acknowledges support from the TESS mission via subaward s3449 from MIT. This work makes use of observations from the LCOGT network. Part of the LCOGT telescope time was granted by NOIRLab through the Mid-Scale Innovations Program (MSIP). MSIP is funded by NSF. This paper is based on observations made with the Las Cumbres Observatory’s education network telescopes that were upgraded through generous support from the Gordon and Betty Moore Foundation. This paper is based on observations made with the MuSCAT instruments, developed by the Astrobiology Center (ABC) in Japan, the University of Tokyo, and Las Cumbres Observatory (LCOGT). MuSCAT3 was developed with financial support by JSPS KAKENHI (JP18H05439) and JST PRESTO (JPMJPR1775), and is located at the Faulkes Telescope North on Maui, HI (USA), operated by LCOGT. MuSCAT4 was developed with financial support provided by the Heising-Simons Foundation (grant 2022-3611), JST grant number JPMJCR1761, and the ABC in Japan, and is located at the Faulkes Telescope South at Siding Spring Observatory (Australia), operated by LCOGT. This research has made use of the Exoplanet Follow-up Observation Program (ExoFOP; DOI: 10.26134/ExoFOP5) website, which is operated by the California Institute of Technology, under contract with the National Aeronautics and Space Administration under the Exoplanet Exploration Program.

SG1 activities at Pine Mountain Observatory are supported in part by the Heising-Simons Foundation and the Roundhouse Foundation.

This work is partly supported by JSPS KAKENHI Grant Numbers JP24H00017, JP24H00248, JP24K00689, JP24K17082, JP24K17083, JSPS Grant-in-Aid for JSPS Fellows Grant Number JP24KJ0241, and JSPS Bilateral Program Number JPJSBP120249910.
This article is based on observations made with the MuSCAT2 instrument, developed by ABC, at Telescopio Carlos Sánchez operated on the island of Tenerife by the IAC in the Spanish Observatorio del Teide.

This paper is based on observations made with the MuSCAT3/4 instruments, developed by the Astrobiology Center (ABC) in Japan, the University of Tokyo, and Las Cumbres Observatory (LCOGT). MuSCAT3 was developed with financial support by JSPS KAKENHI (JP18H05439) and JST PRESTO (JPMJPR1775), and is located at the Faulkes Telescope North on Maui, HI (USA), operated by LCOGT. MuSCAT4 was developed with financial support provided by the Heising-Simons Foundation (grant 2022-3611), JST grant number JPMJCR1761, and the ABC in Japan, and is located at the Faulkes Telescope South at Siding Spring Observatory (Australia), operated by LCOGT.

We acknowledge the use of public TESS data from pipelines at the TESS Science Office and at the TESS Science Processing Operations Center.

This research has made use of the Exoplanet Follow-up Observation Program website, which is operated by the California Institute of Technology, under contract with the National Aeronautics and Space Administration under the Exoplanet Exploration Program.

Resources supporting this work were provided by the NASA High-End Computing (HEC) Program through the NASA Advanced Supercomputing (NAS) Division at Ames Research Center for the production of the SPOC data products.

This paper includes data collected by the TESS mission that are publicly available from the Mikulski Archive for Space Telescopes (MAST).

We acknowledge financial support from the Agencia Estatal de Investigaci\'on of the Ministerio de Ciencia e Innovaci\'on MCIN/AEI/10.13039/501100011033 and the ERDF “A way of making Europe” through project PID2021-125627OB-C32, and from the Centre of Excellence “Severo Ochoa” award to the Instituto de Astrofisica de Canarias.

This paper is based on observations made with observatory time provided to Boyce Research Initiatives and Education Foundation by the Las Cumbres Observatory through its Global Sky Partners program.

We thank T\"urkiye National Observatories for the partial support in using T100 telescope with the project numbers 22BT100-1958 and 23BT100-2045

This work has made use of data from the European Space Agency (ESA) mission Gaia (https://www.cosmos.esa.int/gaia), processed by the Gaia Data Processing and Analysis Consortium (DPAC, https://www.cosmos.esa.int/web/gaia/dpac/consortium).

This research made use of \texttt{exoplanet} \citep{exoplanet:joss,
exoplanet:zenodo} and its dependencies \citep{celerite2,
exoplanet:agol20, exoplanet:arviz,
exoplanet:astropy13, exoplanet:astropy18, exoplanet:kipping13,
exoplanet:luger18}.

\section{Data Availability}
\label{availability}
 The \tess{} data products from which we derived our light curves are publicly available online from the Mikulski Archive for Space Telescopes (MAST). Ground-based observations mentioned in this paper is available from ExoFOP-TESS. Additional data not supplied in the paper will be shared on reasonable request to the lead author.

\bibliographystyle{mnras}
\bibliography{ref}

\clearpage
\newpage

{ 
\itshape \footnotesize \noindent
$^{1}$ University of Southern Queensland, West St, Darling Heights, Toowoomba, Queensland, 4350, Australia \\ 
$^{2}$ Centre for Computational Astrophysics, Flatiron Institute, 162 Fifth Avenue, New York, NY 10010, USA \\
$^{3}$ Department of Astronomy, University of Illinois at Urbana-Champaign, Urbana, IL 61801, USA \\
$^{4}$ Center for Astrophysics \textbar \ Harvard \& Smithsonian, 60 Garden Street, Cambridge, MA 02138, USA \\
$^{5}$ NASA Ames Research Center, Moffett Field, CA 94035, USA \\
$^{6}$Department of Physics and Kavli Institute for Astrophysics and Space Research, Massachusetts Institute of Technology, 77 Massachusetts Ave, Cambridge, MA 02139, USA \\
$^{7}$ Department of Earth, Atmospheric, and Planetary Sciences, Massachusetts Institute of Technology, Cambridge, MA 02139, USA \\
$^{8}$ Department of Aeronautics and Astronautics, Massachusetts Institute of Technology, Cambridge, MA 02139, USA \\
$^{9}$ Department of Astrophysical Sciences, Princeton University, Princeton, NJ 08544, USA \\
$^{10}$ Pine Mountain Observatory, Institute for Fundamental Science, Department of Physics, University of Oregon, Eugene, OR 97403, USA \\
$^{11}$ Komaba Institute for Science, The University of Tokyo, 3-8-1 Komaba, Meguro, Tokyo 153-8902, Japan \\
$^{12}$ Instituto de Astrof\'{i}sica de Canarias (IAC), 38205 La Laguna, Tenerife, Spain \\
$^{13}$ Physics Department, Austin College, Sherman, TX 75090, USA \\ 
$^{14}$ Gruppo Astrofili Palidoro, Fiumicino, Italy \\
$^{15}$ Departamento de Astronom\'{\i}a y Astrof\'{\i}sica, Universidad de Valencia, E-46100 Burjassot, Valencia, Spain \\
$^{16}$ Observatorio Astron\'omico, Universidad de Valencia, E-46980 Paterna, Valencia, Spain \\
$^{17}$ Department of Astronomy, Tsinghua University, Beijing 100084, People's Republic of China \\
$^{18}$ Department of Physics, Novosibirsk State University, Novosibirsk 630090, Russia \\
$^{19}$ Department of Multi-Disciplinary Sciences, Graduate School of Arts and Sciences, The University of Tokyo, 3-8-1 Komaba, Meguro, Tokyo 153-8902, Japan \\
$^{20}$ Hamilton College, 198 College Hill Rd, Clinton, NY 13413, USA \\
$^{21}$ National Research Council Canada, Herzberg Astronomy \& Astrophysics Research Centre, 5071 West Saanich Road, Victoria, BC V9E 2E7, Canada \\
$^{22}$ Astrobiology Center, 2-21-1 Osawa, Mitaka, Tokyo 181-8588, Japan \\
$^{23}$ National Astronomical Observatory of Japan, 2-21-1 Osawa, Mitaka, Tokyo 181-8588, Japan \\
$^{24}$ Department of Electronics, Electrical Engineering and Microelectronics, Silesian University of Technology, Gliwice, Poland \\
$^{25}$ Brierfield Observatory, Bowral, NSW, Australia \\
$^{26}$ AAVSO \\
$^{27}$ Hazelwood Observatory, Australia \\
$^{28}$ CAS Key Laboratory of Optical Astronomy, National Astronomical Observatories, Chinese Academy of Sciences, Beijing 100101, China \\
$^{29}$ Department of Physics and Astronomy, Union College, 807 Union St., Schenectady, NY 12308, USA \\
$^{30}$ Obserwatorium Astronomiczne Niedźwiady, Szubin, Poland \\
$^{31}$ Boyce Research Initiatives and Education Foundation, San Diego, CA, USA \\ 
$^{32}$ Departamento de Astrof\'isica, Universidad de La Laguna (ULL), 38206 La Laguna, Tenerife, Spain \\
$^{33}$ Ankara University, Faculty of Science, Astronomy \& Space Sciences Department, Tandogan, TR-06100, Ankara, T\"urkiye \\
$^{34}$ Ankara University, Astronomy and Space Sciences Research and Application Center (Kreiken Observatory), {İ}ncek Blvd., TR-06837, Ahlatl{\i}bel, Ankara, T\"urkiye \\
}

\appendix
\section{Transit modelling results}

\onecolumn
\begin{landscape}
\fontsize{6}{8}\selectfont
\renewcommand*{\arraystretch}{2}
\begin{longtable}{ccccccccccccc}
\caption{Table of stellar and planetary properties for the final warm sub-Saturn sample. \textbf{Uncertainties are given as the 68\% quantiles unless specified otherwise.} Column descriptions: (1) \tess{} Input Catalog ID, TIC ID; (2) \tess{} Object of Interest name, TOI; (3) Alternate Name, If Applicable (4) stellar radius in solar radius, $R_{\star}$; (5) stellar density in mean solar density, $\rho_{\star}$; (6) inferred stellar density from \tess{} light curves, Fitted $\rho_{\mathrm{circ}}$; (7) planet-to-star radius ratio, $\frac{R_p}{R_{\star}}$; (8) planet radius in Earth radii, $R_p$; (9) impact parameter, $b$; (10) orbital period in days, $P$; (11) time of conjunction, $T_c$; (12) the mode and 68\% \textbf{Highest Density Interval} of the uniformly derived eccentricity inferred from \tess{} light curves, $e$ (13)  the mode and 68\% \textbf{Highest Density Interval} of the uniformly derived argument of periastron from \tess{} light curves, $\omega$ } \\
\toprule
TIC ID & TOI & Alternate Name & $R_{\star}$ $(R_{\odot})$ & $\rho_{\star}$ $(\rho_{\odot})$ & Fitted $\rho_{\mathrm{circ}}$ $(\rho_{\odot})$ & $\frac{R_p}{R_{\star}}$ & $R_p$ $(R_{\oplus})$ & $b$ & $P$ (days) & $T_c$ (BJD - 2457000) & $e$ & $\omega$ $(^\circ)$ \\
\midrule
29781292 & 282.01 & HD 28109 c & $1.397_{-0.032}^{+0.032}$ & $0.511_{-0.082}^{+0.082}$ & $0.24_{-0.14}^{+0.22}$ & $0.0273_{-0.0011}^{+0.0017}$ & $4.23_{-0.25}^{+0.23}$ & $0.62_{-0.40}^{+0.19}$ & $56.00072_{-0.00016}^{+0.00016}$ & $1337.2530_{-0.0039}^{+0.0039}$ & $0.06_{-0.06}^{+0.39}$ & $-50_{-110}^{+20}$ \\
32498058 & 4347.01 &  & $1.591_{-0.043}^{+0.043}$ & $0.397_{-0.056}^{+0.056}$ & $0.45_{-0.27}^{+0.18}$ & $0.0239_{-0.0019}^{+0.0020}$ & $4.16_{-0.33}^{+0.37}$ & $0.43_{-0.29}^{+0.31}$ & $35.35748_{-0.00072}^{+0.00072}$ & $1449.127_{-0.012}^{+0.012}$ & $0.04_{-0.04}^{+0.31}$ & $-160_{-20}^{+200}$ \\
37770169 & 470.01 & TOI-470 b & $0.823_{-0.026}^{+0.026}$ & $1.56_{-0.20}^{+0.20}$ & $1.87_{-0.87}^{+0.72}$ & $0.0475_{-0.0016}^{+0.0017}$ & $4.27_{-0.20}^{+0.19}$ & $0.47_{-0.31}^{+0.23}$ & $12.191491_{-0.000033}^{+0.000033}$ & $1474.4939_{-0.0015}^{+0.0015}$ & $0.04_{-0.04}^{+0.28}$ & $-20_{-20}^{+200}$ \\
47601197 & 2350.01 &  & $1.035_{-0.032}^{+0.032}$ & $0.96_{-0.12}^{+0.12}$ & $1.3_{-0.9}^{+1.8}$ & $0.0570_{-0.0057}^{+0.0074}$ & $6.57_{-0.78}^{+0.72}$ & $0.70_{-0.43}^{+0.18}$ & $8.6197_{-0.0038}^{+0.0038}$ & $1476.5884_{-0.0026}^{+0.0026}$ & $0.17_{-0.16}^{+0.30}$ & $-170_{-130}^{+350}$ \\
48018596 & 2295.01 &  TOI-2295 b & $1.436_{-0.040}^{+0.040}$ & $0.400_{-0.068}^{+0.068}$ & $7.6_{-7.2}^{+8.4}$ & $0.030_{-0.003}^{+0.033}$ & $4.7_{-0.3}^{+2.5}$ & $0.65_{-0.45}^{+0.35}$ & $30.03326_{-0.00014}^{+0.00014}$ & $1713.4523_{-0.0030}^{+0.0030}$ & $0.85_{-0.18}^{+0.09}$ & $63_{-36}^{+82}$ \\
49428710 & 5174.01 & TOI-5174 b & $1.105_{-0.036}^{+0.036}$ & $0.73_{-0.11}^{+0.11}$ & $0.58_{-0.30}^{+0.18}$ & $0.0393_{-0.0017}^{+0.0020}$ & $4.76_{-0.29}^{+0.26}$ & $0.42_{-0.29}^{+0.28}$ & $12.21947_{-0.00038}^{+0.00038}$ & $2527.2968_{-0.0022}^{+0.0022}$ & $0.04_{-0.04}^{+0.28}$ & $-20_{-150}^{+10}$ \\
55525572 & 813.01 & TOI-813 b & $1.864_{-0.037}^{+0.037}$ & $0.183_{-0.020}^{+0.020}$ & $0.216_{-0.057}^{+0.027}$ & $0.03331_{-0.00075}^{+0.00090}$ & $6.80_{-0.21}^{+0.21}$ & $0.29_{-0.20}^{+0.21}$ & $83.89654_{-0.00023}^{+0.00023}$ & $2125.8320_{-0.0022}^{+0.0022}$ & $0.04_{-0.04}^{+0.23}$ & $-10_{-10}^{+190}$ \\
67630845 & 2408.01 &  & $0.752_{-0.020}^{+0.020}$ & $1.87_{-0.19}^{+0.19}$ & $1.66_{-0.97}^{+0.80}$ & $0.0679_{-0.0045}^{+0.0074}$ & $5.73_{-0.51}^{+0.51}$ & $0.47_{-0.32}^{+0.29}$ & $9.46363_{-0.00012}^{+0.00012}$ & $2128.2970_{-0.0020}^{+0.0020}$ & $0.06_{-0.06}^{+0.28}$ & $-150_{-30}^{+150}$ \\
68035559 & 2484.01 &  & $0.875_{-0.025}^{+0.025}$ & $1.32_{-0.16}^{+0.16}$ & $2.0_{-1.0}^{+0.9}$ & $0.0562_{-0.0028}^{+0.0034}$ & $5.41_{-0.34}^{+0.34}$ & $0.48_{-0.32}^{+0.25}$ & $8.327311_{-0.000078}^{+0.000078}$ & $2686.1143_{-0.0014}^{+0.0014}$ & $0.03_{-0.03}^{+0.35}$ & $-170_{-160}^{+350}$ \\
86273567 & 3858.01 &  & $1.138_{-0.036}^{+0.036}$ & $0.70_{-0.11}^{+0.11}$ & $0.86_{-0.37}^{+0.20}$ & $0.0431_{-0.0022}^{+0.0024}$ & $5.37_{-0.35}^{+0.31}$ & $0.37_{-0.25}^{+0.27}$ & $16.59687_{-0.00016}^{+0.00016}$ & $2606.4873_{-0.0021}^{+0.0021}$ & $0.00_{-0.00}^{+0.32}$ & $-170_{-140}^{+350}$ \\
94986319 & 421.01 & TOI-421 c & $0.828_{-0.027}^{+0.027}$ & $1.53_{-0.19}^{+0.19}$ & $3.1_{-1.0}^{+2.3}$ & $0.0519_{-0.0019}^{+0.0013}$ & $4.65_{-0.20}^{+0.22}$ & $0.63_{-0.32}^{+0.11}$ & $16.067536_{-0.000017}^{+0.000017}$ & $2195.30731_{-0.00058}^{+0.00058}$ & $0.30_{-0.24}^{+0.22}$ & $0_{-0}^{+170}$ \\
95079513 & 5504.01 &  & $1.179_{-0.028}^{+0.028}$ & $0.649_{-0.089}^{+0.089}$ & $0.56_{-0.40}^{+0.35}$ & $0.0392_{-0.0028}^{+0.0051}$ & $5.24_{-0.53}^{+0.52}$ & $0.54_{-0.36}^{+0.30}$ & $13.2435_{-0.0019}^{+0.0019}$ & $2563.0517_{-0.0040}^{+0.0040}$ & $0.01_{-0.01}^{+0.35}$ & $-150_{-30}^{+160}$ \\
96317115 & 4806.01 &  & $1.132_{-0.029}^{+0.029}$ & $0.653_{-0.093}^{+0.093}$ & $0.59_{-0.38}^{+0.37}$ & $0.0434_{-0.0030}^{+0.0037}$ & $5.41_{-0.44}^{+0.41}$ & $0.50_{-0.34}^{+0.29}$ & $8.44769_{-0.00014}^{+0.00014}$ & $2248.2923_{-0.0042}^{+0.0042}$ & $0.03_{-0.03}^{+0.34}$ & $-30_{-150}^{+40}$ \\
115010361 & 5163.01 & K2-353 b & $0.985_{-0.028}^{+0.028}$ & $0.94_{-0.13}^{+0.13}$ & $1.5_{-1.0}^{+0.8}$ & $0.0469_{-0.0038}^{+0.0056}$ & $5.16_{-0.57}^{+0.45}$ & $0.48_{-0.33}^{+0.33}$ & $15.46587_{-0.00017}^{+0.00017}$ & $2674.5800_{-0.0049}^{+0.0049}$ & $0.07_{-0.07}^{+0.34}$ & $-10_{-10}^{+190}$ \\
120255950 & 4584.01 & Kepler-1656 b & $0.993_{-0.027}^{+0.027}$ & $0.94_{-0.14}^{+0.14}$ & $14.8_{-8.6}^{+6.3}$ & $0.0442_{-0.0020}^{+0.0023}$ & $4.82_{-0.27}^{+0.26}$ & $0.49_{-0.33}^{+0.28}$ & $31.57875_{-0.00017}^{+0.00017}$ & $2399.6390_{-0.0014}^{+0.0014}$ & $0.79_{-0.12}^{+0.08}$ & $77_{-42}^{+74}$ \\
123846039 & 883.01 &  & $0.975_{-0.025}^{+0.025}$ & $0.98_{-0.14}^{+0.14}$ & $0.92_{-0.25}^{+0.40}$ & $0.0641_{-0.0023}^{+0.0019}$ & $6.79_{-0.28}^{+0.29}$ & $0.53_{-0.28}^{+0.12}$ & $10.057739_{-0.000026}^{+0.000026}$ & $2210.74363_{-0.00045}^{+0.00045}$ & $0.01_{-0.01}^{+0.28}$ & $-160_{-20}^{+170}$ \\
126322519 & 5502.01 &  & $1.349_{-0.047}^{+0.047}$ & $0.442_{-0.059}^{+0.059}$ & $1.38_{-0.81}^{+0.54}$ & $0.0294_{-0.0018}^{+0.0028}$ & $4.41_{-0.38}^{+0.33}$ & $0.45_{-0.30}^{+0.31}$ & $11.62599_{-0.00018}^{+0.00018}$ & $2578.3595_{-0.0027}^{+0.0027}$ & $0.48_{-0.22}^{+0.21}$ & $30_{-20}^{+130}$ \\
142087638 & 2404.01 &  & $1.219_{-0.028}^{+0.028}$ & $0.65_{-0.10}^{+0.10}$ & $1.18_{-0.45}^{+0.23}$ & $0.0319_{-0.0010}^{+0.0013}$ & $4.27_{-0.17}^{+0.19}$ & $0.35_{-0.24}^{+0.25}$ & $20.362773_{-0.000047}^{+0.000047}$ & $2384.8543_{-0.0012}^{+0.0012}$ & $0.24_{-0.16}^{+0.26}$ & $0_{-10}^{+160}$ \\
142276270 & 1136.01 & TOI-1136 d & $0.961_{-0.025}^{+0.025}$ & $1.06_{-0.14}^{+0.14}$ & $0.90_{-0.27}^{+0.25}$ & $0.0447_{-0.0010}^{+0.0011}$ & $4.70_{-0.16}^{+0.16}$ & $0.43_{-0.27}^{+0.17}$ & $12.518598_{-0.000017}^{+0.000017}$ & $1686.06303_{-0.00093}^{+0.00093}$ & $0.01_{-0.01}^{+0.26}$ & $-150_{-20}^{+150}$ \\
146413471 & 6454.01 &  & $0.850_{-0.030}^{+0.030}$ & $1.53_{-0.19}^{+0.19}$ & $0.86_{-0.65}^{+0.42}$ & $0.059_{-0.005}^{+0.013}$ & $6.01_{-0.97}^{+0.70}$ & $0.50_{-0.35}^{+0.36}$ & $22.50116_{-0.00018}^{+0.00018}$ & $2948.8224_{-0.0039}^{+0.0039}$ & $0.07_{-0.07}^{+0.34}$ & $-140_{-20}^{+120}$ \\
148782377 & 1415.01 &  & $1.415_{-0.033}^{+0.033}$ & $0.400_{-0.060}^{+0.060}$ & $0.41_{-0.30}^{+0.23}$ & $0.0316_{-0.0013}^{+0.0022}$ & $4.98_{-0.31}^{+0.27}$ & $0.52_{-0.37}^{+0.31}$ & $14.42071_{-0.00024}^{+0.00024}$ & $1933.1786_{-0.0021}^{+0.0021}$ & $0.03_{-0.03}^{+0.32}$ & $-160_{-20}^{+190}$ \\
154716798 & 1239.01 &  & $0.775_{-0.025}^{+0.025}$ & $1.80_{-0.21}^{+0.21}$ & $1.88_{-0.58}^{+0.35}$ & $0.0546_{-0.0013}^{+0.0018}$ & $4.65_{-0.20}^{+0.19}$ & $0.36_{-0.24}^{+0.22}$ & $12.639716_{-0.000013}^{+0.000013}$ & $2604.09601_{-0.00050}^{+0.00050}$ & $0.02_{-0.02}^{+0.24}$ & $-20_{-160}^{+40}$ \\
158978373 & 823.01 &  & $1.315_{-0.037}^{+0.037}$ & $0.485_{-0.062}^{+0.062}$ & $0.46_{-0.17}^{+0.70}$ & $0.0479_{-0.0031}^{+0.0019}$ & $6.76_{-0.40}^{+0.42}$ & $0.78_{-0.28}^{+0.07}$ & $13.533764_{-0.000027}^{+0.000027}$ & $2339.8325_{-0.0014}^{+0.0014}$ & $0.13_{-0.13}^{+0.31}$ & $-30_{-150}^{+70}$ \\
159781361 & 2019.01 & TOI-2019 b & $1.771_{-0.046}^{+0.046}$ & $0.189_{-0.030}^{+0.030}$ & $0.18_{-0.11}^{+0.08}$ & $0.0310_{-0.0015}^{+0.0022}$ & $6.08_{-0.40}^{+0.38}$ & $0.48_{-0.32}^{+0.28}$ & $15.34683_{-0.00042}^{+0.00042}$ & $3431.584_{-0.040}^{+0.040}$ & $0.05_{-0.05}^{+0.29}$ & $-20_{-160}^{+40}$ \\
169191182 & 6414.01 &  & $1.302_{-0.038}^{+0.038}$ & $0.538_{-0.063}^{+0.063}$ & $0.67_{-0.37}^{+0.26}$ & $0.0394_{-0.0030}^{+0.0036}$ & $5.66_{-0.53}^{+0.46}$ & $0.46_{-0.31}^{+0.28}$ & $17.76079_{-0.00019}^{+0.00019}$ & $2987.7320_{-0.0033}^{+0.0033}$ & $0.04_{-0.04}^{+0.31}$ & $-170_{-130}^{+350}$ \\
176780257 & 4187.01 &  & $2.245_{-0.060}^{+0.060}$ & $0.110_{-0.014}^{+0.014}$ & $0.124_{-0.088}^{+0.050}$ & $0.0185_{-0.0011}^{+0.0016}$ & $4.61_{-0.35}^{+0.35}$ & $0.47_{-0.32}^{+0.35}$ & $30.8803_{-0.0010}^{+0.0010}$ & $1425.263_{-0.023}^{+0.023}$ & $0.03_{-0.03}^{+0.31}$ & $-160_{-20}^{+210}$ \\
193220940 & 2806.01 &  & $1.419_{-0.037}^{+0.037}$ & $0.451_{-0.049}^{+0.049}$ & $0.61_{-0.37}^{+0.28}$ & $0.0301_{-0.0018}^{+0.0023}$ & $4.71_{-0.36}^{+0.31}$ & $0.44_{-0.30}^{+0.32}$ & $11.40462_{-0.00012}^{+0.00012}$ & $2298.6216_{-0.0066}^{+0.0066}$ & $0.09_{-0.09}^{+0.27}$ & $-10_{-20}^{+190}$ \\
200296137 & 4303.01 &  & $1.768_{-0.050}^{+0.050}$ & $0.248_{-0.033}^{+0.033}$ & $2.2_{-1.5}^{+1.1}$ & $0.0223_{-0.0017}^{+0.0029}$ & $4.46_{-0.46}^{+0.42}$ & $0.48_{-0.33}^{+0.33}$ & $8.610948_{-0.000054}^{+0.000054}$ & $1438.4919_{-0.0037}^{+0.0037}$ & $0.69_{-0.10}^{+0.16}$ & $92_{-62}^{+56}$ \\
200723869 & 257.01 & TOI-257 b & $1.968_{-0.061}^{+0.061}$ & $0.179_{-0.031}^{+0.031}$ & $0.38_{-0.11}^{+0.09}$ & $0.0336_{-0.0010}^{+0.0010}$ & $7.22_{-0.31}^{+0.31}$ & $0.38_{-0.24}^{+0.18}$ & $18.387689_{-0.000031}^{+0.000031}$ & $2121.26902_{-0.00089}^{+0.00089}$ & $0.28_{-0.11}^{+0.27}$ & $160_{-140}^{+10}$ \\
224299081 & 5619.01 &  & $1.353_{-0.033}^{+0.033}$ & $0.404_{-0.072}^{+0.072}$ & $0.94_{-0.43}^{+0.22}$ & $0.0509_{-0.0021}^{+0.0027}$ & $7.56_{-0.40}^{+0.39}$ & $0.38_{-0.26}^{+0.29}$ & $37.99842_{-0.00039}^{+0.00039}$ & $2626.9022_{-0.0025}^{+0.0025}$ & $0.31_{-0.15}^{+0.27}$ & $30_{-20}^{+140}$ \\
230386259 & 1785.01 &  & $2.873_{-0.091}^{+0.091}$ & $0.0711_{-0.0084}^{+0.0084}$ & $0.20_{-0.13}^{+0.10}$ & $0.0182_{-0.0009}^{+0.0018}$ & $5.89_{-0.48}^{+0.44}$ & $0.52_{-0.35}^{+0.28}$ & $11.00239_{-0.00013}^{+0.00013}$ & $2944.8951_{-0.0034}^{+0.0034}$ & $0.38_{-0.18}^{+0.27}$ & $160_{-140}^{+10}$ \\
232608943 & 4600.01 & TOI-4600 b & $0.826_{-0.023}^{+0.023}$ & $1.56_{-0.16}^{+0.16}$ & $1.23_{-0.33}^{+0.25}$ & $0.0741_{-0.0017}^{+0.0018}$ & $6.69_{-0.25}^{+0.23}$ & $0.37_{-0.23}^{+0.18}$ & $82.68808_{-0.00050}^{+0.00050}$ & $2419.3921_{-0.0023}^{+0.0023}$ & $0.01_{-0.01}^{+0.29}$ & $-150_{-10}^{+130}$ \\
235943205 & 2264.01 &  & $0.982_{-0.024}^{+0.024}$ & $1.09_{-0.11}^{+0.11}$ & $0.8_{-0.4}^{+1.6}$ & $0.0461_{-0.0038}^{+0.0034}$ & $4.92_{-0.39}^{+0.41}$ & $0.77_{-0.38}^{+0.09}$ & $121.33438_{-0.00061}^{+0.00061}$ & $1827.0277_{-0.0034}^{+0.0034}$ & $0.01_{-0.01}^{+0.45}$ & $-140_{-40}^{+160}$ \\
237015185 & 5604.01 &  & $1.442_{-0.038}^{+0.038}$ & $0.444_{-0.049}^{+0.049}$ & $0.71_{-0.14}^{+0.06}$ & $0.0477_{-0.0016}^{+0.0015}$ & $7.50_{-0.32}^{+0.31}$ & $0.22_{-0.15}^{+0.21}$ & $11.180142_{-0.000053}^{+0.000053}$ & $2840.57595_{-0.00091}^{+0.00091}$ & $0.20_{-0.13}^{+0.23}$ & $170_{-160}^{+10}$ \\
237071229 & 2039.01 &  & $1.491_{-0.041}^{+0.041}$ & $0.389_{-0.046}^{+0.046}$ & $0.88_{-0.56}^{+0.48}$ & $0.0355_{-0.0017}^{+0.0034}$ & $5.97_{-0.48}^{+0.42}$ & $0.53_{-0.36}^{+0.28}$ & $11.248032_{-0.000075}^{+0.000075}$ & $3359.8230_{-0.0038}^{+0.0038}$ & $0.34_{-0.24}^{+0.24}$ & $-180_{-170}^{+340}$ \\
237929468 & 4940.01 &  & $1.167_{-0.030}^{+0.030}$ & $0.593_{-0.068}^{+0.068}$ & $0.3_{-0.2}^{+1.1}$ & $0.0547_{-0.0084}^{+0.0095}$ & $7.1_{-1.2}^{+1.1}$ & $0.87_{-0.38}^{+0.06}$ & $25.86788_{-0.00013}^{+0.00013}$ & $2317.0395_{-0.0035}^{+0.0035}$ & $0.17_{-0.17}^{+0.38}$ & $-130_{-50}^{+160}$ \\
238854767 & 4653.01 &  & $1.273_{-0.037}^{+0.037}$ & $0.614_{-0.066}^{+0.066}$ & $0.58_{-0.33}^{+0.19}$ & $0.0292_{-0.0017}^{+0.0018}$ & $4.06_{-0.26}^{+0.26}$ & $0.43_{-0.29}^{+0.31}$ & $16.37610_{-0.00018}^{+0.00018}$ & $2080.370_{-0.011}^{+0.011}$ & $0.03_{-0.03}^{+0.28}$ & $-20_{-160}^{+30}$ \\
243244680 & 5065.01 & EPIC 211945201 b & $1.385_{-0.036}^{+0.036}$ & $0.412_{-0.047}^{+0.047}$ & $1.1_{-0.7}^{+1.2}$ & $0.0397_{-0.0036}^{+0.0051}$ & $6.15_{-0.69}^{+0.62}$ & $0.66_{-0.41}^{+0.20}$ & $19.492148_{-0.000046}^{+0.000046}$ & $2575.8510_{-0.0014}^{+0.0014}$ & $0.48_{-0.38}^{+0.15}$ & $20_{-10}^{+160}$ \\
259230140 & 4384.01 &  & $1.683_{-0.048}^{+0.048}$ & $0.324_{-0.044}^{+0.044}$ & $0.52_{-0.10}^{+0.04}$ & $0.03680_{-0.00085}^{+0.00083}$ & $6.76_{-0.28}^{+0.22}$ & $0.24_{-0.17}^{+0.19}$ & $14.317047_{-0.000075}^{+0.000075}$ & $3070.24892_{-0.00082}^{+0.00082}$ & $0.17_{-0.10}^{+0.27}$ & $180_{-160}^{+0}$ \\
260367888 & 3354.01 &  & $1.651_{-0.045}^{+0.045}$ & $0.272_{-0.034}^{+0.034}$ & $0.86_{-0.51}^{+0.35}$ & $0.0334_{-0.0017}^{+0.0020}$ & $6.05_{-0.39}^{+0.35}$ & $0.48_{-0.32}^{+0.28}$ & $46.38061_{-0.00031}^{+0.00031}$ & $2983.7508_{-0.0028}^{+0.0028}$ & $0.45_{-0.21}^{+0.21}$ & $130_{-120}^{+30}$ \\
261646133 & 6724.01 &  & $1.362_{-0.030}^{+0.030}$ & $0.432_{-0.050}^{+0.050}$ & $0.37_{-0.22}^{+0.14}$ & $0.0317_{-0.0015}^{+0.0018}$ & $4.74_{-0.25}^{+0.27}$ & $0.44_{-0.30}^{+0.31}$ & $42.86122_{-0.00024}^{+0.00024}$ & $3117.8039_{-0.0051}^{+0.0051}$ & $0.03_{-0.03}^{+0.29}$ & $-150_{-30}^{+150}$ \\
269838674 & 5413.01 &  & $1.313_{-0.062}^{+0.062}$ & $0.48_{-0.14}^{+0.14}$ & $1.06_{-0.62}^{+0.45}$ & $0.0416_{-0.0026}^{+0.0038}$ & $6.08_{-0.57}^{+0.48}$ & $0.47_{-0.31}^{+0.29}$ & $12.6122_{-0.0015}^{+0.0015}$ & $2534.9205_{-0.0030}^{+0.0030}$ & $0.33_{-0.24}^{+0.24}$ & $160_{-160}^{+10}$ \\
272167060 & 3495.01 &  & $1.932_{-0.055}^{+0.055}$ & $0.196_{-0.029}^{+0.029}$ & $1.8_{-1.2}^{+0.9}$ & $0.0190_{-0.0013}^{+0.0017}$ & $4.07_{-0.35}^{+0.33}$ & $0.49_{-0.33}^{+0.32}$ & $8.413073_{-0.000055}^{+0.000055}$ & $2347.1613_{-0.0041}^{+0.0041}$ & $0.68_{-0.10}^{+0.16}$ & $59_{-30}^{+86}$ \\
272670038 & 6079.01 &  & $1.458_{-0.040}^{+0.040}$ & $0.375_{-0.071}^{+0.071}$ & $0.59_{-0.21}^{+0.81}$ & $0.0507_{-0.0043}^{+0.0036}$ & $7.99_{-0.67}^{+0.66}$ & $0.88_{-0.13}^{+0.04}$ & $104.29005_{-0.00079}^{+0.00079}$ & $1912.4553_{-0.0056}^{+0.0056}$ & $0.15_{-0.15}^{+0.41}$ & $-150_{-90}^{+330}$ \\
280031353 & 2300.02 &  & $0.788_{-0.021}^{+0.021}$ & $1.59_{-0.21}^{+0.21}$ & $1.58_{-0.41}^{+0.44}$ & $0.0638_{-0.0019}^{+0.0019}$ & $5.49_{-0.22}^{+0.22}$ & $0.45_{-0.27}^{+0.15}$ & $15.443278_{-0.000038}^{+0.000038}$ & $2646.96070_{-0.00054}^{+0.00054}$ & $0.00_{-0.00}^{+0.26}$ & $-20_{-160}^{+30}$ \\
286864983 & 772.01 &  & $0.803_{-0.030}^{+0.030}$ & $1.62_{-0.22}^{+0.22}$ & $0.86_{-0.14}^{+0.31}$ & $0.0789_{-0.0026}^{+0.0015}$ & $6.84_{-0.31}^{+0.30}$ & $0.65_{-0.14}^{+0.05}$ & $11.0163622_{-0.0000086}^{+0.0000086}$ & $1575.94382_{-0.00072}^{+0.00072}$ & $0.25_{-0.24}^{+0.18}$ & $-40_{-110}^{+0}$ \\
288636342 & 1692.01 &  & $1.503_{-0.037}^{+0.037}$ & $0.430_{-0.046}^{+0.046}$ & $0.33_{-0.16}^{+0.07}$ & $0.0338_{-0.0011}^{+0.0014}$ & $5.58_{-0.22}^{+0.26}$ & $0.37_{-0.25}^{+0.31}$ & $17.728702_{-0.000090}^{+0.000090}$ & $2725.0142_{-0.0015}^{+0.0015}$ & $0.00_{-0.00}^{+0.31}$ & $-30_{-140}^{+10}$ \\
288636342 & 1692.02 &  & $1.503_{-0.037}^{+0.037}$ & $0.430_{-0.046}^{+0.046}$ & $0.40_{-0.11}^{+0.10}$ & $0.0487_{-0.0010}^{+0.0011}$ & $8.00_{-0.27}^{+0.25}$ & $0.41_{-0.24}^{+0.17}$ & $32.20798_{-0.00013}^{+0.00013}$ & $2717.4853_{-0.0011}^{+0.0011}$ & $0.01_{-0.01}^{+0.25}$ & $-20_{-160}^{+20}$ \\
299158887 & 1554.01 & Kepler-63 b & $0.894_{-0.028}^{+0.028}$ & $1.35_{-0.15}^{+0.15}$ & $1.47_{-0.32}^{+0.67}$ & $0.0613_{-0.0016}^{+0.0013}$ & $5.96_{-0.24}^{+0.22}$ & $0.67_{-0.14}^{+0.06}$ & $9.434148_{-0.000011}^{+0.000011}$ & $2444.89600_{-0.00046}^{+0.00046}$ & $0.00_{-0.00}^{+0.31}$ & $-160_{-20}^{+200}$ \\
308994098 & 790.01 &  & $1.467_{-0.033}^{+0.033}$ & $0.452_{-0.077}^{+0.077}$ & $0.117_{-0.019}^{+0.022}$ & $0.0464_{-0.0024}^{+0.0026}$ & $7.44_{-0.43}^{+0.41}$ & $0.897_{-0.016}^{+0.014}$ & $199.5761_{-0.0014}^{+0.0014}$ & $1352.4918_{-0.0040}^{+0.0040}$ & $0.46_{-0.19}^{+0.23}$ & $-52_{-75}^{+3}$ \\
328012209 & 2309.01 &  & $1.315_{-0.031}^{+0.031}$ & $0.489_{-0.058}^{+0.058}$ & $0.31_{-0.21}^{+0.16}$ & $0.0281_{-0.0015}^{+0.0024}$ & $4.11_{-0.33}^{+0.27}$ & $0.51_{-0.34}^{+0.31}$ & $20.26563_{-0.00013}^{+0.00013}$ & $2044.3335_{-0.0045}^{+0.0045}$ & $0.06_{-0.06}^{+0.32}$ & $-40_{-120}^{+20}$ \\
330690135 & 5170.01 & K2-333 b & $1.203_{-0.034}^{+0.034}$ & $0.688_{-0.068}^{+0.068}$ & $0.68_{-0.27}^{+0.15}$ & $0.0458_{-0.0018}^{+0.0019}$ & $6.02_{-0.29}^{+0.29}$ & $0.36_{-0.25}^{+0.26}$ & $14.7571_{-0.0014}^{+0.0014}$ & $2558.6192_{-0.0022}^{+0.0022}$ & $0.02_{-0.02}^{+0.26}$ & $-160_{-20}^{+170}$ \\
332564140 & 5400.01 &  & $0.932_{-0.026}^{+0.026}$ & $1.20_{-0.16}^{+0.16}$ & $1.15_{-0.63}^{+0.32}$ & $0.0431_{-0.0017}^{+0.0024}$ & $4.42_{-0.27}^{+0.22}$ & $0.39_{-0.27}^{+0.32}$ & $19.80666_{-0.00014}^{+0.00014}$ & $2970.5304_{-0.0021}^{+0.0021}$ & $0.00_{-0.00}^{+0.31}$ & $-20_{-160}^{+30}$ \\
332660150 & 938.01 &  & $0.990_{-0.025}^{+0.025}$ & $0.93_{-0.12}^{+0.12}$ & $0.32_{-0.24}^{+0.28}$ & $0.0456_{-0.0035}^{+0.0063}$ & $5.11_{-0.57}^{+0.49}$ & $0.61_{-0.40}^{+0.27}$ & $8.80735_{-0.00014}^{+0.00014}$ & $2193.8016_{-0.0034}^{+0.0034}$ & $0.34_{-0.29}^{+0.21}$ & $-128_{-15}^{+96}$ \\
336126253 & 5193.01 &  & $0.875_{-0.023}^{+0.023}$ & $1.42_{-0.14}^{+0.14}$ & $1.42_{-0.71}^{+0.52}$ & $0.0581_{-0.0028}^{+0.0028}$ & $5.55_{-0.29}^{+0.31}$ & $0.44_{-0.29}^{+0.27}$ & $18.99711_{-0.00011}^{+0.00011}$ & $2408.1536_{-0.0025}^{+0.0025}$ & $0.01_{-0.01}^{+0.29}$ & $-20_{-160}^{+40}$ \\
346855283 & 6646.01 &  & $1.407_{-0.042}^{+0.042}$ & $0.500_{-0.064}^{+0.064}$ & $0.45_{-0.23}^{+0.13}$ & $0.0357_{-0.0021}^{+0.0024}$ & $5.51_{-0.39}^{+0.38}$ & $0.41_{-0.28}^{+0.29}$ & $17.75135_{-0.00011}^{+0.00011}$ & $2763.6574_{-0.0038}^{+0.0038}$ & $0.04_{-0.04}^{+0.25}$ & $-150_{-30}^{+160}$ \\
347357977 & 2370.01 &  & $1.039_{-0.028}^{+0.028}$ & $0.888_{-0.087}^{+0.087}$ & $0.94_{-0.47}^{+0.27}$ & $0.0480_{-0.0020}^{+0.0026}$ & $5.49_{-0.29}^{+0.30}$ & $0.40_{-0.27}^{+0.29}$ & $12.682573_{-0.000070}^{+0.000070}$ & $2262.5120_{-0.0034}^{+0.0034}$ & $0.00_{-0.00}^{+0.29}$ & $-20_{-160}^{+50}$ \\
350020859 & 5141.01 & HD 89345 b & $1.768_{-0.041}^{+0.041}$ & $0.190_{-0.027}^{+0.027}$ & $0.425_{-0.098}^{+0.037}$ & $0.03619_{-0.00090}^{+0.00090}$ & $6.98_{-0.24}^{+0.23}$ & $0.24_{-0.17}^{+0.22}$ & $11.81427_{-0.00072}^{+0.00072}$ & $2556.00490_{-0.00079}^{+0.00079}$ & $0.30_{-0.11}^{+0.23}$ & $30_{-20}^{+130}$ \\
352682207 & 4010.02 & TOI-4010 d & $0.794_{-0.037}^{+0.037}$ & $1.73_{-0.29}^{+0.29}$ & $1.23_{-0.61}^{+0.98}$ & $0.0668_{-0.0032}^{+0.0037}$ & $5.82_{-0.40}^{+0.40}$ & $0.62_{-0.35}^{+0.16}$ & $14.70901_{-0.00024}^{+0.00024}$ & $2721.4265_{-0.0019}^{+0.0019}$ & $0.03_{-0.03}^{+0.34}$ & $-150_{-30}^{+140}$ \\
356472238 & 5382.01 &  & $1.735_{-0.043}^{+0.043}$ & $0.248_{-0.027}^{+0.027}$ & $0.81_{-0.60}^{+0.58}$ & $0.0260_{-0.0019}^{+0.0035}$ & $5.12_{-0.56}^{+0.47}$ & $0.53_{-0.36}^{+0.33}$ & $14.5425_{-0.0012}^{+0.0012}$ & $2961.9538_{-0.0056}^{+0.0056}$ & $0.45_{-0.24}^{+0.25}$ & $20_{-20}^{+140}$ \\
358248442 & 2419.01 &  & $1.383_{-0.073}^{+0.073}$ & $0.40_{-0.11}^{+0.11}$ & $0.31_{-0.15}^{+0.11}$ & $0.03241_{-0.00097}^{+0.00098}$ & $4.90_{-0.30}^{+0.29}$ & $0.46_{-0.31}^{+0.25}$ & $18.881126_{-0.000071}^{+0.000071}$ & $3169.7765_{-0.0034}^{+0.0034}$ & $0.03_{-0.03}^{+0.31}$ & $-150_{-30}^{+140}$ \\
366410512 & 5101.01 & K2-98 b & $1.567_{-0.045}^{+0.045}$ & $0.307_{-0.043}^{+0.043}$ & $0.26_{-0.20}^{+0.21}$ & $0.0320_{-0.0031}^{+0.0047}$ & $5.63_{-0.69}^{+0.65}$ & $0.53_{-0.36}^{+0.34}$ & $10.136709_{-0.000066}^{+0.000066}$ & $3247.8284_{-0.0097}^{+0.0097}$ & $0.13_{-0.13}^{+0.28}$ & $-150_{-30}^{+170}$ \\
373017346 & 4450.01 &  & $1.133_{-0.028}^{+0.028}$ & $0.68_{-0.11}^{+0.11}$ & $1.08_{-0.54}^{+0.39}$ & $0.0335_{-0.0012}^{+0.0016}$ & $4.17_{-0.21}^{+0.19}$ & $0.46_{-0.31}^{+0.25}$ & $10.692506_{-0.000050}^{+0.000050}$ & $2759.8776_{-0.0011}^{+0.0011}$ & $0.14_{-0.14}^{+0.26}$ & $-170_{-160}^{+340}$ \\
382200986 & 4409.01 &  & $0.731_{-0.017}^{+0.017}$ & $2.01_{-0.19}^{+0.19}$ & $2.11_{-0.31}^{+0.15}$ & $0.0959_{-0.0016}^{+0.0017}$ & $7.67_{-0.22}^{+0.22}$ & $0.20_{-0.14}^{+0.18}$ & $92.49187_{-0.00028}^{+0.00028}$ & $2076.3477_{-0.0012}^{+0.0012}$ & $0.00_{-0.00}^{+0.21}$ & $-170_{-10}^{+200}$ \\
384984325 & 6109.01 &  & $0.987_{-0.030}^{+0.030}$ & $1.05_{-0.14}^{+0.14}$ & $2.9_{-1.3}^{+0.6}$ & $0.0425_{-0.0018}^{+0.0020}$ & $4.60_{-0.27}^{+0.23}$ & $0.35_{-0.24}^{+0.29}$ & $8.538723_{-0.000018}^{+0.000018}$ & $1798.0622_{-0.0017}^{+0.0017}$ & $0.38_{-0.15}^{+0.22}$ & $30_{-10}^{+130}$ \\
408114363 & 2450.01 &  & $0.778_{-0.026}^{+0.026}$ & $1.83_{-0.21}^{+0.21}$ & $1.31_{-0.41}^{+0.29}$ & $0.0763_{-0.0021}^{+0.0023}$ & $6.49_{-0.28}^{+0.30}$ & $0.38_{-0.25}^{+0.21}$ & $17.88518_{-0.00016}^{+0.00016}$ & $2179.1757_{-0.0011}^{+0.0011}$ & $0.03_{-0.03}^{+0.30}$ & $-150_{-10}^{+130}$ \\
408137826 & 941.01 &  & $1.182_{-0.041}^{+0.041}$ & $0.61_{-0.10}^{+0.10}$ & $0.78_{-0.28}^{+0.16}$ & $0.0534_{-0.0019}^{+0.0020}$ & $6.90_{-0.35}^{+0.35}$ & $0.36_{-0.24}^{+0.24}$ & $8.514604_{-0.000080}^{+0.000080}$ & $2198.2771_{-0.0012}^{+0.0012}$ & $0.00_{-0.00}^{+0.31}$ & $-170_{-150}^{+350}$ \\
410214986 & 200.01 & DS Tuc A b & $0.941_{-0.030}^{+0.030}$ & $1.02_{-0.17}^{+0.17}$ & $1.40_{-0.25}^{+0.28}$ & $0.0479_{-0.0038}^{+0.0038}$ & $4.92_{-0.42}^{+0.42}$ & $0.41_{-0.19}^{+0.11}$ & $8.1382230_{-0.0000030}^{+0.0000030}$ & $2040.33358_{-0.00031}^{+0.00031}$ & $0.10_{-0.10}^{+0.24}$ & $0_{-20}^{+160}$ \\
418301643 & 6247.01 &  & $1.502_{-0.050}^{+0.050}$ & $0.329_{-0.040}^{+0.040}$ & $0.60_{-0.32}^{+0.20}$ & $0.0263_{-0.0013}^{+0.0015}$ & $4.33_{-0.25}^{+0.28}$ & $0.44_{-0.30}^{+0.28}$ & $24.091699_{-0.000099}^{+0.000099}$ & $1771.2414_{-0.0037}^{+0.0037}$ & $0.26_{-0.22}^{+0.21}$ & $0_{-10}^{+160}$ \\
427320001 & 2112.01 &  & $1.604_{-0.047}^{+0.047}$ & $0.352_{-0.045}^{+0.045}$ & $0.39_{-0.11}^{+0.04}$ & $0.02578_{-0.00082}^{+0.00089}$ & $4.52_{-0.21}^{+0.19}$ & $0.28_{-0.19}^{+0.23}$ & $14.01113_{-0.00020}^{+0.00020}$ & $2729.8128_{-0.0021}^{+0.0021}$ & $0.02_{-0.02}^{+0.24}$ & $-160_{-130}^{+340}$ \\
428703516 & 5761.01 &  & $0.729_{-0.026}^{+0.026}$ & $2.10_{-0.27}^{+0.27}$ & $2.52_{-0.88}^{+0.49}$ & $0.0660_{-0.0023}^{+0.0023}$ & $5.25_{-0.27}^{+0.25}$ & $0.34_{-0.23}^{+0.25}$ & $19.542220_{-0.000083}^{+0.000083}$ & $2722.3125_{-0.0013}^{+0.0013}$ & $0.02_{-0.02}^{+0.27}$ & $-10_{-20}^{+190}$ \\
432280671 & 6012.01 &  & $1.203_{-0.034}^{+0.034}$ & $0.592_{-0.086}^{+0.086}$ & $0.65_{-0.23}^{+0.10}$ & $0.0576_{-0.0012}^{+0.0013}$ & $7.57_{-0.30}^{+0.24}$ & $0.32_{-0.23}^{+0.26}$ & $16.2114_{-0.0021}^{+0.0021}$ & $2861.6587_{-0.0016}^{+0.0016}$ & $0.00_{-0.00}^{+0.27}$ & $-10_{-170}^{+50}$ \\
437756191 & 6203.01 &  & $1.328_{-0.033}^{+0.033}$ & $0.423_{-0.072}^{+0.072}$ & $0.93_{-0.39}^{+0.22}$ & $0.0502_{-0.0018}^{+0.0023}$ & $7.34_{-0.40}^{+0.32}$ & $0.38_{-0.25}^{+0.26}$ & $25.1738_{-0.0036}^{+0.0036}$ & $2855.6080_{-0.0023}^{+0.0023}$ & $0.30_{-0.16}^{+0.25}$ & $170_{-150}^{+10}$ \\
441590110 & 5187.01 &  & $1.313_{-0.032}^{+0.032}$ & $0.484_{-0.061}^{+0.061}$ & $0.85_{-0.42}^{+0.24}$ & $0.0476_{-0.0030}^{+0.0032}$ & $6.82_{-0.45}^{+0.50}$ & $0.39_{-0.26}^{+0.30}$ & $17.85980_{-0.00019}^{+0.00019}$ & $2596.5073_{-0.0038}^{+0.0038}$ & $0.19_{-0.19}^{+0.22}$ & $170_{-170}^{+10}$ \\
445805961 & 1710.01 & TOI-1710 b & $0.953_{-0.023}^{+0.023}$ & $1.16_{-0.18}^{+0.18}$ & $1.09_{-0.14}^{+0.06}$ & $0.04915_{-0.00052}^{+0.00058}$ & $5.12_{-0.14}^{+0.12}$ & $0.18_{-0.13}^{+0.17}$ & $24.283374_{-0.000023}^{+0.000023}$ & $2905.43142_{-0.00053}^{+0.00053}$ & $0.02_{-0.02}^{+0.22}$ & $-20_{-150}^{+20}$ \\
468979441 & 5493.01 &  & $1.047_{-0.029}^{+0.029}$ & $0.92_{-0.10}^{+0.10}$ & $0.67_{-0.28}^{+0.23}$ & $0.0479_{-0.0014}^{+0.0017}$ & $5.50_{-0.23}^{+0.24}$ & $0.46_{-0.31}^{+0.22}$ & $24.43835_{-0.00028}^{+0.00028}$ & $1502.459_{-0.013}^{+0.013}$ & $0.06_{-0.06}^{+0.27}$ & $-30_{-130}^{+10}$ \\
\bottomrule
\end{longtable}

\label{tab:final_table}
\end{landscape}

\begin{figure*}
\centering
    \includegraphics[width=\textwidth]{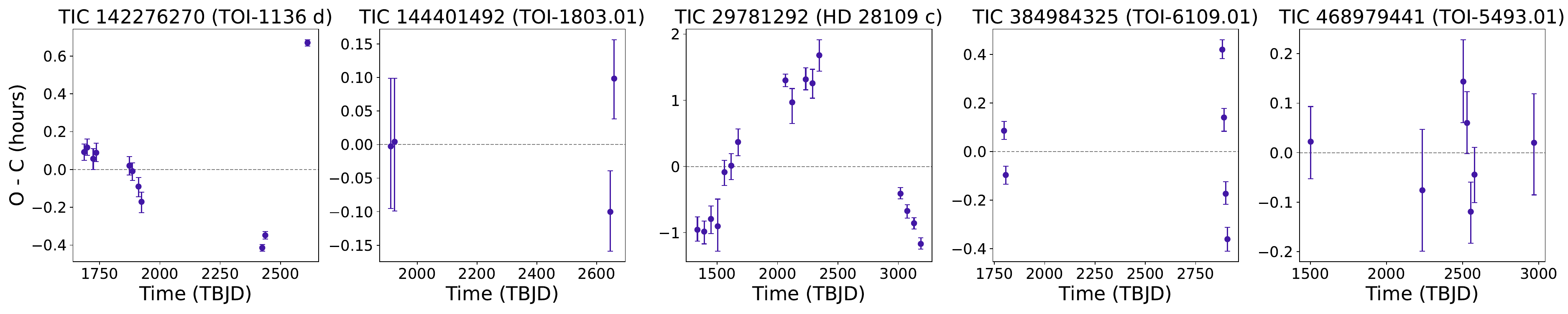}
\caption{The O-C diagram (Observed-Calculated)
of mid-transit times versus time in BJD-2457000 for the candidates that we fitted transit timing variations. We include a horizontal dashed line centered at zero in each panel for
reference.}
\label{fig:ttvplots}
\end{figure*}

\twocolumn
\begin{figure*}
\centering
\includegraphics[width=\textwidth]{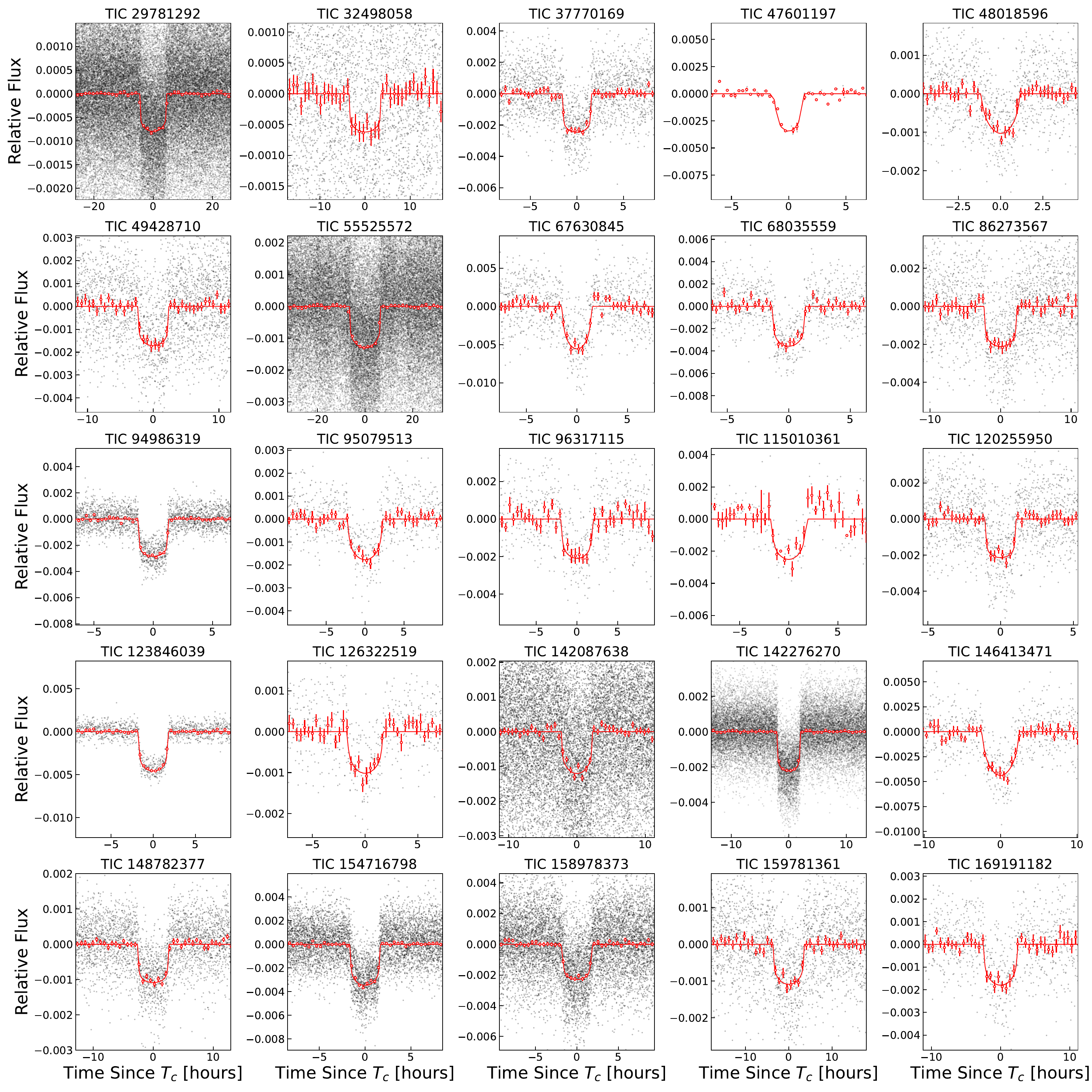}
\caption{Phase-folded light curves for all our sub-Saturns, over-plotted with the best-fitted transit models. For planets with significant transit timings, the phases are calculated respecting to the best fit $T_c$ of each transit.}
\label{fig:lightcurves}
\end{figure*}
\twocolumn
\begin{figure*}
\centering
\includegraphics[width=\textwidth]{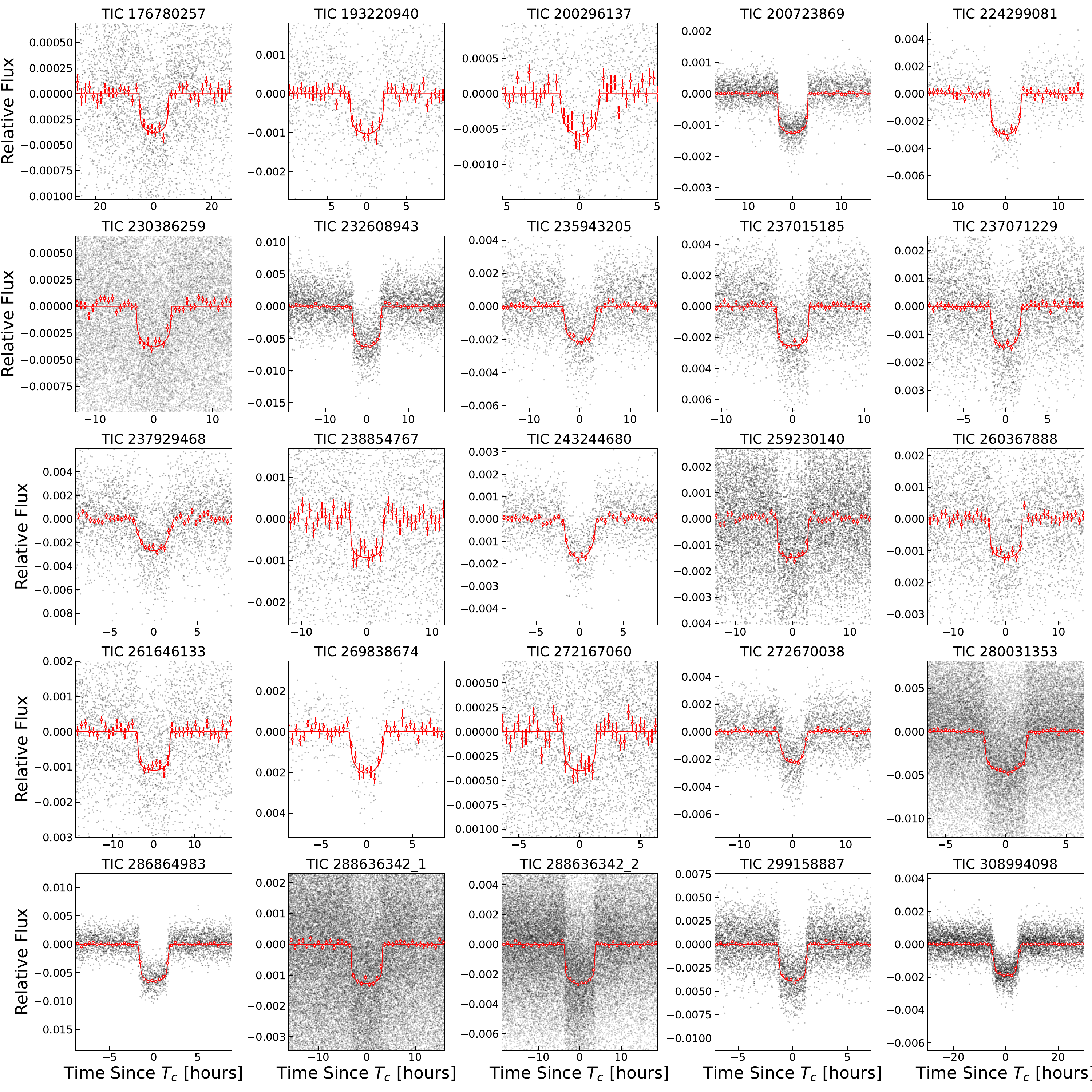}
\caption{Phase-folded light curves for all our sub-Saturns, over-plotted with the best-fitted transit models. For planets with significant transit timings, the phases are calculated respecting to the best fit $T_c$ of each transit.}
\end{figure*}
\twocolumn
\begin{figure*}
\centering
\includegraphics[width=\textwidth]{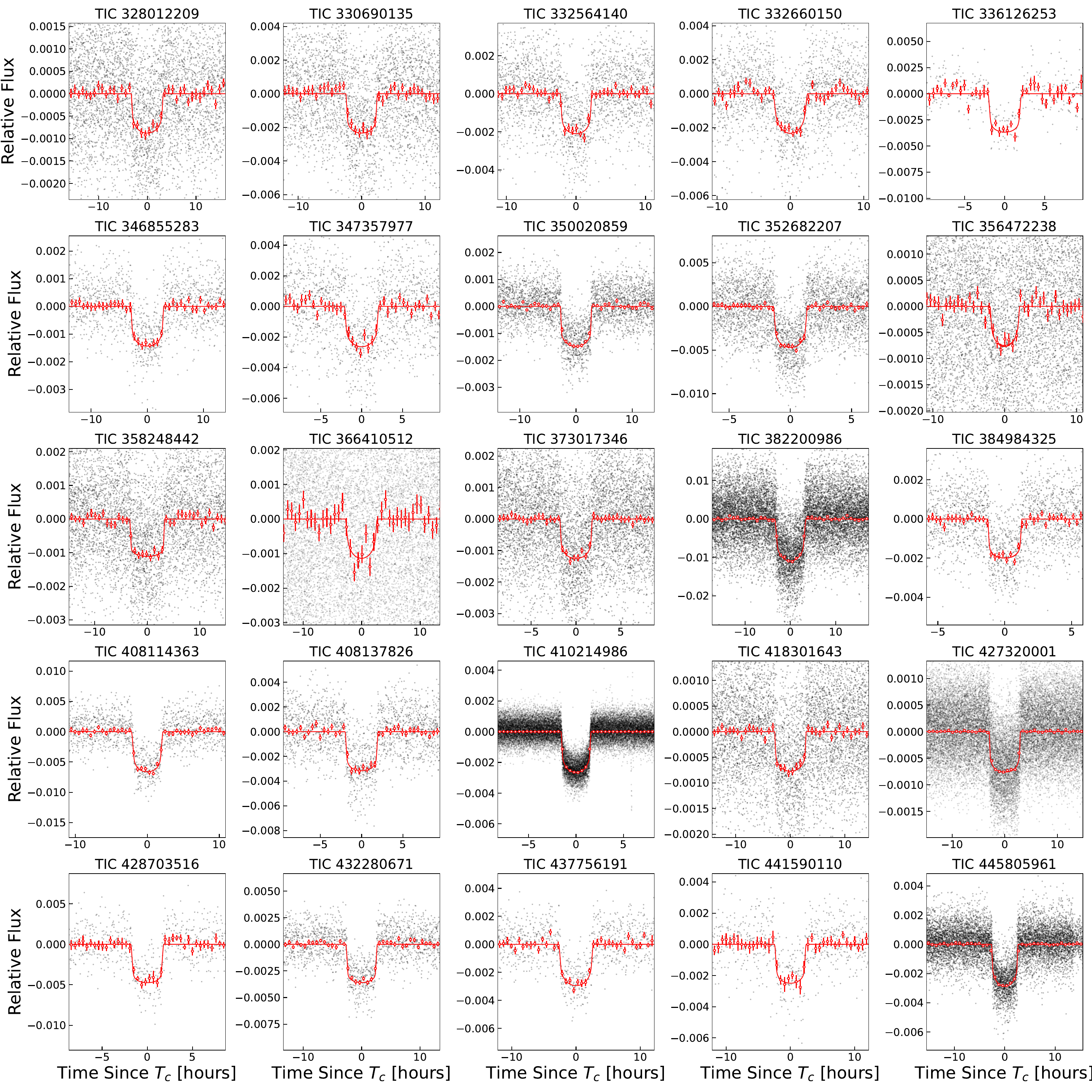}
\caption{Phase-folded light curves for all our sub-Saturns, over-plotted with the best-fitted transit models. For planets with significant transit timings, the phases are calculated respecting to the best fit $T_c$ of each transit.}
\end{figure*}
\twocolumn
\begin{figure*}
\centering
\includegraphics[width=\textwidth]{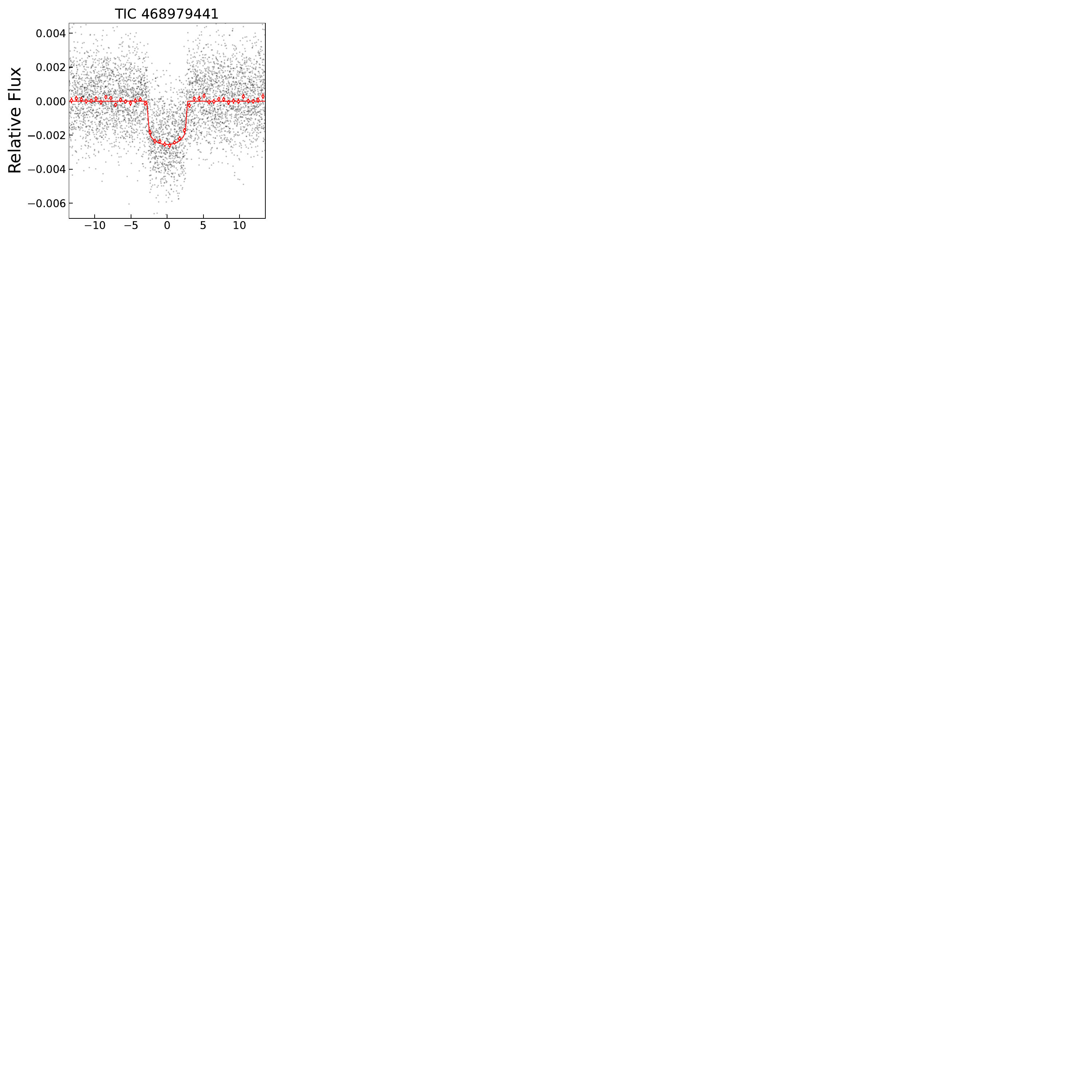}
\caption{Phase-folded light curves for all our sub-Saturns, over-plotted with the best-fitted transit models. For planets with significant transit timings, the phases are calculated respecting to the best fit $T_c$ of each transit.}
\end{figure*}

\bsp	
\label{lastpage}
\end{document}